\documentclass[aps,amsmath,showpacs,amsfonts,10pt]{revtex4}
\usepackage{epsfig,graphicx}
\usepackage[english]{babel}
\usepackage{amsfonts}
\usepackage{amsmath}
\usepackage{latexsym}
\usepackage{graphics,bm}
\usepackage{natbib}
\usepackage{dcolumn}
\usepackage{bm}
\usepackage{rotating}

\begin{document}

\title{Achieving the Quantum Ground State of a Mechanical Oscillator using a Bose-Einstein Condensate with Back-Action and Cold Damping feedback schemes}

\author{ Sonam Mahajan$^{1}$ , Neha Aggarwal$^{1,2}$, Aranya B Bhattacherjee$^{2}$ and ManMohan$^{1}$}

\address{$^{1}$Department of Physics and Astrophysics, University of Delhi, Delhi-110007, India} \address{$^{2}$Department of Physics, ARSD College, University of Delhi (South Campus), New Delhi-110021, India}

\begin{abstract}
We present a detailed study to show the possibility of approaching the quantum ground-state of a hybrid optomechanical quantum device formed by a Bose-Einstein condensate (BEC) confined inside a high-finesse optical cavity with an oscillatory end mirror. Cooling is achieved using two experimentally realizable schemes: back-action cooling and cold damping quantum feedback cooling. In both the schemes, we found that increasing the two body atom-atom interaction brings the mechanical oscillator to its quantum ground state. It has been observed that back-action cooling is more effective in the good cavity limit while the cold damping cooling scheme is more relevant in the bad cavity limit. It is also shown that in the cold damping scheme, the device is more efficient in the presence of BEC than in the absence of BEC.
\end{abstract}


\maketitle

\section{Introduction}
Optomechanical cooling of micro- and nano-mechanical resonators to their quantum ground state, is used in a wide variety of sensitive measurements such as detection of weak forces \citep{1,2}, small masses \citep{3} and small displacements \citep{4}. Such mechanical resonators can also be used as tools for quantum metrology \citep{5,6} or as a medium to couple hybrid quantum systems \citep{7,8}. In recent years, different types of optomechanical systems have established considerable cooling of the vibrational modes of mechanical resonators interacting with optical cavity \citep{9,10}. In such systems, the cavity exerts radiation pressure on the mirror. The dynamics of the mirror is coherently controlled by an external pump laser. Therefore, by modifying the pump power, a strong coupling regime can be easily acheived \citep{11}. In such a strong coupling regime, the mechanical oscillator can be cooled to its quantum ground state using the dynamics of back-action \citep{12,13}. Experimentally, significant cooling of the mechanical mode of the resonator coupled to an optical cavity has been attained by using two different ways of radiation pressure interaction between the intracavity field and a vibrational mode : back-action cooling \citep{9,14,15} and cold damping feedback \citep{10,16,17}. Self cooling of the mechanical resonator via dynamical back-action arises due to its interaction with the optical cavity through radiation pressure \citep{9,14,15}. Depending upon the laser detuning, the correlations, induced by the cavity delay, between the Brownian motion of the resonator and the radiation pressure force lead to either cooling or amplification. Experimentally, a single mechanical mode has been cooled using these effects \citep{18,19,20}. With the back-action cooling or self cooling, there is self modification in the dynamics of the mechanical system as the off resonant operation of the cavity results in the retarded dynamical back-action on the system \citep{9,18,19,20,21}. Cold damping quantum feedback technique is used to cool the oscillating mirror by applying a viscous force on the mirror via radiation pressure generated by another intensity modulated laser beam on the back of the mirror \citep{22,23,24,25}. This technique increases the damping of the system without any additional noise \citep{26,27}. Other feedback scheme requires very large mechanical quality factor \citep{folman} for the ground state cooling of the oscillator whereas cooling can be much more conveniently achieved in the cold damping scheme. The optomechanical systems composed of ultracold atoms enclose within a cavity has grabbed much attention\citep{28,29,30,31,32,33,34,35,36,37,38,39,40}. The resonance frequency of the cavity is altered due to the strong interaction of the condensed atoms with the cavity mode \citep{41}. An all-optical transistor based on a coupled Bose-Einstein condensate(BEC) cavity system \citep{30,42} has been proposed \citep{43}. An optomechanical system consisting of BEC in an optical cavity has been investigated recently \citep{pater}. By virtue of interaction of BEC with mechanical oscillator, such hybrid system was helpful in acheiving state engineering of the mechanical mode of oscillator, which can be easily controlled and highy insensitive to noise effects. Experimentally, interaction between the ultracold atoms and vibrating membrane has been studied \citep{56}. In our previous work, we have investigated how the stochastic cooling feedback technique together with a gas of condensate confined in an optical cavity can be used to detect weak forces and coherently controls the sensitivity of this hybrid optomechanical system \citep{sonam}.

Motivated by these interesting features in the field of cavity opto-mechanics and BEC, we compare the two schemes i.e., back-action cooling and cold damping quantum feedback, for cooling an optomechanical system in which a BEC is coupled to an optical cavity with a movable mirror. By comparing these two schemes for such a system, we found that the cold damping quantum feedback scheme cools the mechanical oscillator to its ground state in the bad cavity limit ($\kappa>>\omega_{m}$) i.e, when cavity bandwidth is greater than the mechanical frequency while the back-action cooling is efficient in the opposite limit of good cavity ($\kappa<<\omega_{m}$) i.e., when cavity bandwidth is smaller than the mechanical frequency. We also show that for both the schemes, in the presence of atom - atom coupling, the interaction of the vibrational mode of the mirror, single mode of the intracavity field and the condensate fluctuations give rise to normal mode splitting.
   
\section{Model Hamiltonian}

In this section, we introduce our model and describe the Hamiltonian for our system. Our model consists of an elongated cigar shaped gas of N two-level BEC atoms of $^{87} Rb$ in the $|F=1>$ state with mass $m$ and transition frequency $\omega_{a}$ of the  $|F=1>$ $\rightarrow$ $|F'=2>$ transition of the $D_{2}$ line. The BEC atoms strongly interact with a single one-dimensional quantized cavity mode of frequency $\omega_{c}$. The cavity has one mirror fixed and another mirror movable of mass $M_{m}$ which acts as a mechanical oscillator that is free to oscillate at mechanical frequency $\omega_{m}$. The coherent laser field with frequency $\omega_{p}$ drives the system through the fixed cavity mirror with amplitude $\eta'$. The cavity and the laser field are treated quantum mechanically as the standing wave and the pump laser are weak (small photon number). The single-particle Hamiltonian in the rotating wave and dipole approximation for the system under consideration is read as \citep{28}:

\begin{equation}
H_{0} = H_{a}+H_{c}+H_{m}+H_{ca}+H_{cm},
\end{equation}

where $H_{a}=\frac{p^{2}}{2m}$ represents the kinetic energy of the condensate. $H_{c}$ is given by $- \hbar \Delta_{c} b^{\dagger}b-i\hbar \eta^{'} (b-b^{\dagger})$ in which the first term gives the energy of the light mode, with lowering (raising) operator $b$ ($b^{\dagger}$) ($[b,b^{\dagger}]=1$) and cavity-pump detuning $\Delta_{c}(=\omega_{p}-\omega_{c})$. The second term in $H_{c}$ is the energy due to external pump laser. A band pass filter is used in the detection scheme \citep{44} such that only a single vibrational mode can be considered by neglecting the several other mechanical degrees of freedom arising from the radiation pressure. $H_{m}=\hbar \omega_{m} c^{\dagger}c$ gives the energy of the single vibrational mode of the cantilever with annihilation (creation) operator $c$ ($c^{\dagger}$) ($[c,c^{\dagger}]=1$). As a result of the pressure exerted by the intracavity photons on the mirrors, there is an optomechanical coupling between the vibrating mirror and the cavity field. Hence a force is exerted on the movable mirror by the electromagnetic field which depends on the number of photons in the cavity. This field is phase shifted by $2 k l_{m}$ where $k$ is the wave vector of light field and  $l_{m}$ is the displacement of the mirror from its equilibrium position. The scattering of the photons into other modes of the electromagnetic field can be ignored and hence validating the single mode approximation of the cavity field. Also, a reservoir having finite temperature $T$ is attached to the movable mirror. In the absence of the radiation pressure exerted by the optical cavity field, the cantilever can undergo Brownian motion due to its contact with thermal environment. $H_{ca}= \cos^2(k^{'}x)(\hbar \upsilon_{0} b^{\dagger}b+V_{cl})$ describes the interaction of the light field with the BEC which involves the classical potential $V_{cl}$, where $\upsilon_{0}$ is the optical lattice barrier height per photon given by $\upsilon_{0}=\frac{g_{0}^{2}}{\Delta_{a}}$ with $g_{0}$ and $\Delta_{a}$($=\omega_{p}-\omega_{a}$) representing the mirror-atom coupling and atom-pump detuning respectively. We take $\upsilon_{0}>0$ as in this case condensate atom moves towards the nodes of the optical field due to which the lowest bound state is localized at these positions. Therefore compared to the case $\upsilon_{0}<0$, the interaction between the optical cavity field and the condensate atom is decreased for $\upsilon_{0}>0$. $H_{cm}=-\hbar \epsilon \omega_{m} b^{\dagger}b(c+c^{\dagger})$ represents the energy due to non linear dispersive coupling between the position quadrature of the movable mirror and the intensity of light field where $\epsilon$ is the mirror-photon coupling.
The Hamiltonian in second quantized form with the two-body interaction is given as :

\begin{equation}
H = \int d\vec{r}\ \Psi^{\dagger}(\vec{r})H_{0}\Psi(\vec{r}) + \frac{4 \pi a_{s}\hbar^{2}}{2m} \int d\vec{r}\ \Psi^{\dagger}(\vec{r})\Psi^{\dagger}(\vec{r})\Psi(\vec{r})\Psi(\vec{r}),
\end{equation}
where $\Psi(\vec{r})$ is the atom field operator and $a_{s}$ is the s-wave scattering length. Using $\Psi(\vec{r})=\sum_{i} a_{i}w(\vec{r}-\vec{r_{i}})$, we derive the Bose-Hubbard Hamiltonian. Here, $w(\vec{r}-\vec{r_{i}})$ is the wannier function and $a_{i}$ is the corresponding annihilation operator for the $i_{th}$ bosonic atom. Experimentally, the optical lattice depth is tuned such that time scales over which the experiment is performed is smaller than the time scales over which tunneling takes place, hence the tunneling of the atoms into neighbouring wells can be neglected in deriving the Hamiltonian. Therefore, Hamiltonian reads as :

\begin{eqnarray}\label{hom}\nonumber
H &=& K_{0}\sum_{j}a_{j}^{\dagger}a_{j}- \hbar \Delta_{c} b^{\dagger}b+\hbar \omega_{m} c^{\dagger}c+ P_{0}(\hbar \upsilon_{0}b^{\dagger}b+V_{cl}) \sum_{j}a_{j}^{\dagger}a_{j}-\hbar \epsilon \omega_{m} b^{\dagger}b(c+c^{\dagger})\nonumber \\ &-& i\hbar \eta^{'} (b-b^{\dagger})+\frac{\upsilon}{2}\sum_{j}a_{j}^{\dagger}a_{j}^{\dagger}a_{j}a_{j},
\end{eqnarray}

where $K_{0}=\int d\vec{r} w(\vec{r}-\vec{r}_{j})\left\lbrace \left( -\dfrac{\hbar^2 \nabla^{2}}{2m}\right) \right\rbrace w(\vec{r}-\vec{r}_{j})$ is the onsite kinetic energy of the atoms. $P_{0}=\int d\vec{r} w(\vec{r}-\vec{r}_{j}) \cos^2(k^{'}x)w(\vec{r}-\vec{r}_{j})$ is the effective onsite potential energy of the atoms. The last term in the above Hamiltonian represents the two-body atom-atom coupling where $\upsilon=\dfrac{4\pi a_{s}\hbar^{2}}{m}\int d\vec{r}|w(\vec{r})|^{4}$ represents the effective onsite atom-atom interaction energy.

\section{Linearization of Quantum Langevin Equations}

Now, we study the quantum langevin equations(QLEs) of the system. The QLEs of motion for the boson field operator $a_{j}$, cavity photons $b$ and movable mirror (mode) operator $c$ are given as:

\begin{equation}\label{b}
\dot{a_{j}}(t)=-i\frac{K_{0}}{\hbar}a_{j}(t)-i\frac{P_{0}}{\hbar}(\hbar \upsilon_{0}b^{\dagger}(t) b(t)+V_{cl})a_{j}(t)-i\frac{\upsilon}{\hbar}a_{j}^{\dagger}(t)a_{j}(t)a_{j}(t),
\end{equation}

\begin{equation}\label{a}
\dot{b}(t)=-i P_{0}\upsilon_{0}b(t) \sum_{j} a_{j}^{\dagger}(t) a_{j}(t)+i \Delta_{c} b(t)+i \epsilon \omega_{m} b(t) [c(t)+c^{\dagger}(t)]+\eta'-\frac{\kappa}{2}b(t)+\sqrt{\kappa}b_{in}(t),
\end{equation}

\begin{equation}\label{am}
\dot{c}(t)=-i \omega_{m} c(t)+i \epsilon \omega_{m} b^{\dagger}(t) b(t)-\Gamma_{m} c(t)+\sqrt{\Gamma_{m}} \xi_{m}(t).
\end{equation}

Due to leakage of photons from the mirror, the cavity field is damped. Therefore, we have introduced $\kappa$ as the cavity field damping rate. Also, the interaction of the cantilever with environment damps the mechanical mode with damping rate $\Gamma_{m}$. During the experiment, the condensate atoms are robust and therefore the depletion of atoms is not significant. Since the condensate temperature is much smaller than $\hbar\kappa/k_{B}$, so, there is negligible coherent amplification of the condensate motion. The mechanical mode is also affected by a random Brownian force which has $\xi$ as zero mean value. The vaccum radiation input noise is represented by $b_{in}(t)$. The correlation functions for the input noise operators are given in the Appendix I. The QLEs are linearized around the steady state as  $b(t)\rightarrow \beta+\delta b(t)$, $c(t)\rightarrow \gamma+\delta c(t)$ and $a_{j}(t)\rightarrow \frac{\sqrt{N}+\delta a(t)}{\sqrt{M}}$ where $\beta\left(=\frac{\eta'}{-i \Delta_{c}+\frac{\kappa}{2}+i P_{0}\upsilon_{0}N-i 2 Re(\gamma)\, \epsilon \, \omega_{m}}\right)$,$\gamma\left(Re(\gamma)= \frac{\epsilon \, \omega_{m}^{2} |\beta|^{2}}{\omega_{m}^{2}+\Gamma_{m}^{2}}\right)$ and $\sqrt{\frac{N}{M}}$ are the steady state values of cavity mode, mechanical mode and condensate density respectively. $N(=\sum_{j} a_{j}^{\dagger}(t) a_{j}(t))$ atoms occupy $M$ number of lattice sites. It is assumed that all the sites of optical lattice are same, therefore $a_{j}(t)$ is replaced by $a(t)$. By introducing the amplitude and phase quadratures as  $\delta q_{a}(t)=[\delta a(t)+ \delta a^{\dagger}(t)]$, $\delta p_{a}(t)=i[\delta a^{\dagger}(t)- \delta a(t)]$, $\delta q_{b}(t)=[\delta b(t)+ \delta b^{\dagger}(t)]$, $\delta p_{b}(t)=i[\delta b^{\dagger}(t)- \delta b(t)]$, $\delta q(t)=[\delta c(t)+ \delta c^{\dagger}(t)]$, $\delta p(t)=i[\delta c^{\dagger}(t)- \delta c(t)]$, $q_{in}(t)=[b_{in}(t)+b_{in}^{\dagger}(t)]$ and $p_{in}(t)=i[b_{in}^{\dagger}(t)-b_{in}(t)]$, the linearized QLEs for the system are given as follows:

\begin{equation}
\delta \dot{q_{a}}(t)=\beta_{1} \delta p_{a}(t),
\end{equation}

\begin{equation}
\delta \dot{p_{a}}(t)=-\beta_{2} \delta q_{a}(t)-2 g_{c}\delta q_{b}(t),
\end{equation}

\begin{equation}
\delta \dot{q_{b}}(t)=-\frac{\kappa}{2} \delta q_{b}(t)+\sqrt{\kappa} q_{in}(t)-\Delta_{d} \delta p_{b}(t),
\end{equation}

\begin{equation}
\delta \dot{p_{b}}(t)=-\frac{\kappa}{2} \delta p_{b}(t)-2 g_{c} \delta q_{a}(t)+2G \beta \delta q(t)+\sqrt{\kappa}p_{in}(t)+\Delta_{d} \delta q_{b}(t),
\end{equation}

\begin{equation}
\delta \dot{q}(t)=\omega_{m} \delta p(t),
\end{equation}

\begin{equation}
\delta \dot{p}(t)=-\omega_{m} \delta q(t)+ 2 G \beta \delta q_{b}(t)-\Gamma_{m} \delta p(t)+W(t),
\end{equation}

where $\Delta_{d}=-P_{0}\upsilon_{0}N+\Delta_{c}+2G \gamma$, $g_{c}=P_{0}\upsilon_{0} \beta \sqrt{N}$, $G=\epsilon \omega_{m}$, $\nu=K_{0}/\hbar+P_{0}\upsilon_{0} \beta^{2}+P_{0}V_{cl}/\hbar$, $U_{eff}= \frac{\upsilon N}{\hbar M}$, $\beta_{1}=\nu+U_{eff}$, $\beta_{2}=\nu+ 3U_{eff}$ and $W(t)=i \sqrt{\Gamma_{m}}[\xi_{m}^{\dagger}(t)-\xi_{m}(t)]$ which satisfies the correlation given in the Appendix I.

Let us now understand how the radiation pressure cools the mechanical oscillator. Radiation pressure exerted on the movable mirror by the optical cavity mode forms a system which acts as another reservoir connected to the mechanical oscillator when the cavity is properly detuned. As a result, the effective temperature of the vibrational mode is the temperature between the two reservoirs i.e., between the initial bath temperature and temperature of this effective optical reservoir. This effective temperature is zero in practice. Hence the quantum ground state is acheived when the coupling to the initial reservoir, given by the damping rate of the movable mirror $\Gamma_{m}$, is much smaller than the coupling to the effective optical reservoir. This explains us that the radiation pressure coupling should be strong for significant cooling of the mechanical oscillator. Now, we shall see in the next sections how the different cooling techniques help in cooling the mechanical oscillator to its quantum ground state.

\section{Back-Action Cooling}

Back-action dynamics have been realized in a diverse variety of physical systems, including those of ultracold atoms \citep{30,31}.The randomness present in the unavoidable stochastic back-action forces is due to the photon shot noise. These forces arise from the radiation pressure.

\begin{figure}[h]
\hspace{-0.0cm}
\includegraphics [scale=0.8]{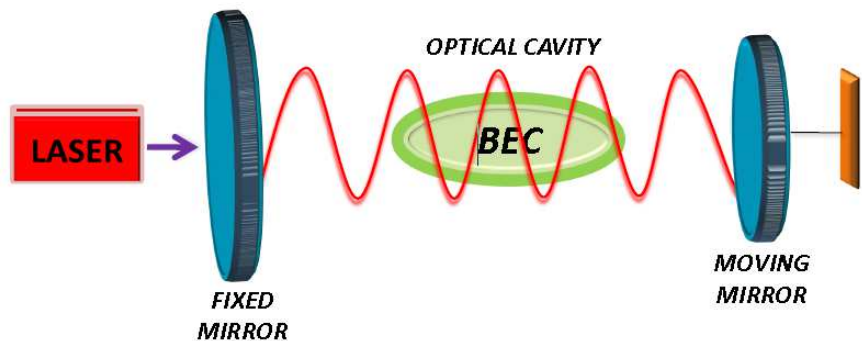}
\caption{(color online) Setup for Back-Action Cooling. It configures an optomechanical system involving Bose Einstein Condensate (BEC) inside a Fabry-Perot cavity with one fixed mirror and another movable light-end mirror. The laser pump is used to drive the cavity mode.}
\label{1}
\end{figure}

In this section, we evaluate effective frequency, effective damping rate and displacement spectrum for the mechanical oscillator (mirror) in the back-action cooling scheme. Also we show how the ground state cooling of the mirror is approached in this scheme.

The displacement spectrum in fourier space is evaluated from 

\begin{equation}
S_{q}(\omega)=\frac{1}{4 \pi} \int d \omega' e^{-i(\omega+\omega')t} \langle \delta q(\omega) \delta q(\omega')+\delta q(\omega') \delta q(\omega) \rangle,
\end{equation}

using the correlations in fourier space given in the Appendix I.
Therefore, the displacement spectrum in the fourier space for the movable mirror is given as :

\begin{equation}
S_{q}(\omega)=\vert \chi_{eff}(\omega)\vert^{2} [S_{th}(\omega) + S_{rp}(\omega , \Delta_{d})],
\end{equation}

where $S_{th}(\omega)$ is the thermal noise spectrum arising from the Brownian motion of the mirror and $S_{rp}(\omega , \Delta_{d})$ is the radiation pressure spectrum including the quantum fluctuations of the condensate. $\chi_{eff}(\omega)$ is the effective susceptibility of the oscillator.

Here

\begin{equation}\label{th}
S_{th}(\omega)=\frac{\Gamma_{m}}{\omega_{m}} \omega\coth \left ( {\frac{\hbar \omega}{2 k_{B} T}} \right ) ,
\end{equation}

\begin{equation}\label{rp}
S_{rp}(\omega , \Delta_{d})=\frac{4G^{2} \beta^{2} \kappa (\omega^{2}-\beta_{1} \beta_{2})^{2} (\Delta_{d}^{2}+\omega^{2}+\frac{\kappa^{2}}{4})}{X(\omega)},
\end{equation}

where
\begin{equation}
 X(\omega)=16g_{c}^{4} \Delta_{d}^2 \beta_{1}^{2}-8g_{c}^{2} \Delta_{d} \beta_{1}(\omega^{2}-\beta_{1} \beta_{2})(\Delta_{d}^{2}+\frac{\kappa^{2}}{4}-\omega^{2})+(\omega^{2}-\beta_{1}\beta_{2})^{2} \left[\omega^{2} \kappa^{2}+(\Delta_{d}^{2}+\frac{\kappa^{2}}{4}-\omega^{2})^{2}\right].
 \end{equation}

\begin{equation}
\chi_{eff}(\omega)=\frac{\omega_{m}}{\left\lbrace(\omega_{m}^{2}-\omega^{2}+i \omega \Gamma_{m})+\chi_{1}(\omega)\right\rbrace},
\end{equation}

where 
\begin{equation}
\chi_{1}(\omega)=\frac{4G^{2}\beta^{2} \Delta_{d} \omega_{m}(\omega^{2}-\beta_{1}\beta_{2})}{\left[(\omega^{2}-\beta_{1}\beta_{2})(\Delta_{d}^{2}+\frac{\kappa^{2}}{4}-\omega^{2}+i \omega \kappa)-4g_{c}^{2}\Delta
_{d}\beta_{1}\right]}.
\end{equation}

$\chi_{eff}(\omega)$ is the effective susceptibility of the resonator altered by the radiation pressure and condensate fluctuations with

\begin{equation}
\vert \chi_{eff}(\omega) \vert^{2}=\frac{\omega_{m}^{2}}{\left[(\omega_{m}^{eff}(\omega)^{2}-\omega^{2})^{2}+\omega^{2} \Gamma_{m}^{eff}(\omega)^{2}\right]}
\end{equation}

\vspace{5 mm}

The effective mechanical susceptibility of the oscillator gives us the effective resonance frequency and effective damping rate as :

\begin{equation}\label{omega}
\omega_{m}^{eff}(\omega)=\left[ \omega_{m}^{2} + \omega_{m}^{op} \right]^{1/2},
\end{equation}

where

\begin{equation}\label{op}
\omega_{m}^{op}=\frac{4G^{2}\beta^{2} \Delta_{d} \omega_{m}(\omega^{2}-\beta_{1}\beta_{2})[(\omega^{2}-\beta_{1}\beta_{2})(\Delta_{d}^{2}+\frac{\kappa^{2}}{4}-\omega^{2})-4g_{c}^{2}\Delta
_{d}\beta_{1}]}{X(\omega)},
\end{equation}

\begin{equation}\label{gamma}
\Gamma_{m}^{eff}(\omega)=\Gamma_{m} - \frac{4G^{2}\beta^{2} \Delta_{d} \omega_{m} \kappa(\omega^{2}-\beta_{1}\beta_{2})^{2}}{X(\omega)}.
\end{equation}

\begin{figure}[h]
\hspace{-0.0cm}
\begin{tabular}{cc}
\includegraphics [scale=0.80]{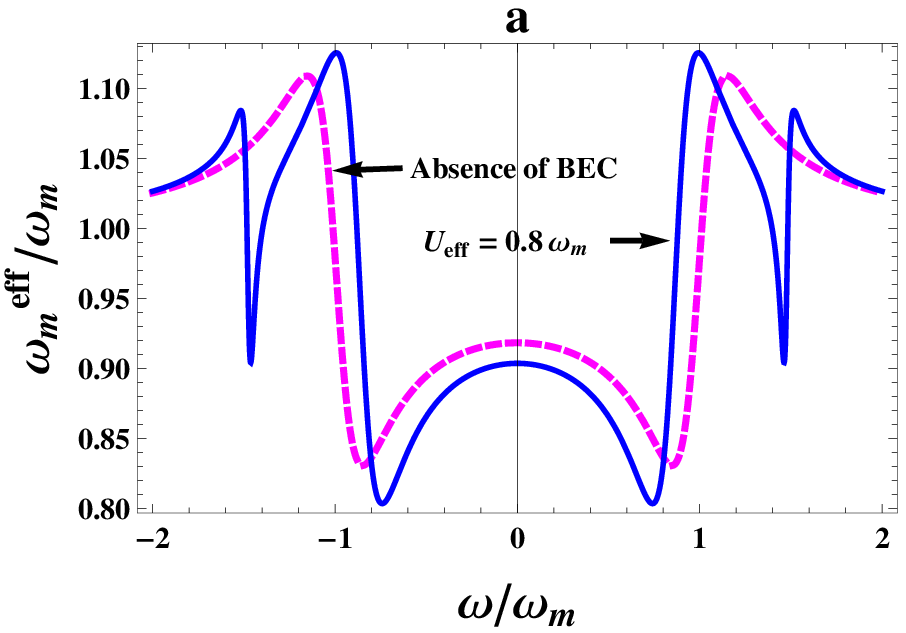}& \includegraphics [scale=0.80] {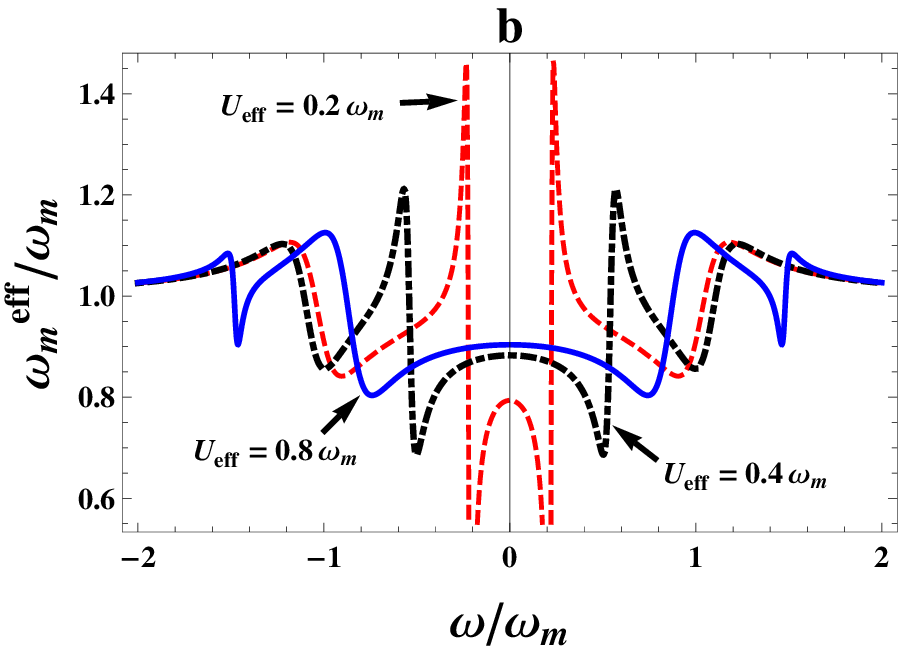}\\
\end{tabular}
\caption{(color online) Plot of normalized effective mechanical frequency ($\omega_{m}^{eff}/\omega_{m}$) as a function of dimensionless frequency in back-action cooling scheme. Fig. 2(a) shows the variation of the dimensionless effective mechanical frequency in the absence of BEC (dashed line) and in the presence of BEC (solid line) with $U_{eff}=0.8\omega_{m}$. Fig. 2(b) represents the dimensionless effective mechanical frequency for three different values of atomic two-body interaction with $U_{eff}=0.2\omega_{m}$ (dashed line), $U_{eff}=0.4\omega_{m}$ (dot dashed line) and $U_{eff}=0.8\omega_{m}$ (solid line). Other parameters used are : $\Gamma_{m}=10^{-5}\omega_{m}$, $\Delta_{d}=-\omega_{m}$, $\kappa=0.3\omega_{m}$, $G=4\omega_{m}$,$\beta=0.05$, $\nu=0.01\omega_{m}$,$g_{c}=0.3\omega_{m}$ and $k_{B}T/\hbar\omega_{m}=10^{3}$.}
\label{2}
\end{figure}

Eqn.(\ref{omega})and(\ref{op}) show that the mechanical frequency of the mirror gets modified by the quantum fluctuations of mirror due to the radiation pressure and condensate fluctuations. This is the so-called optical spring effect. It can be observed that the frequency due to the optical spring term does not get altered significantly for high resonance frequencies, such as those of Refs.\citep{18,19,20} where $\omega_{m} \gtrsim 1$ MHz. 

Experimentally, the mirror may have mechanical frequency varying from $2 \pi \times 100$ Hz \citep{58}, $2 \pi \times 10$ kHz \citep{34} or $2 \pi \times 73.5$ MHz \citep{59} with corresponding damping rate from $2 \pi \times 10^{-3}$ Hz \citep{58}, $2 \pi \times 3.22$ Hz \citep{34} or $2 \pi \times 1.3$ kHz \citep{59}. A high finesse Fabry Perot optical cavity having decay rate $\kappa=2 \pi \times 8.75$ kHz \citep{55} ($2 \pi \times 0.66$ MHz \citep{31}) consists of cloud of BEC with an order of $10^{6}$ $^{87} Rb$ atoms \citep{56} interacting with cavity field may have coherent coupling strength as $g_{0}=2 \pi \times 5.86$ kHz \citep{55} ($2 \pi \times 14.4$ MHz \citep{31}). The loss of photons through the cavity mirrors decreases the energy of the cavity mode which further minimises the interaction of atom light field. This loss of photons can be reduced in high finesse optical cavities.

\begin{figure}[h]
\hspace{-0.0cm}
\begin{tabular}{cc}
\includegraphics [scale=0.80]{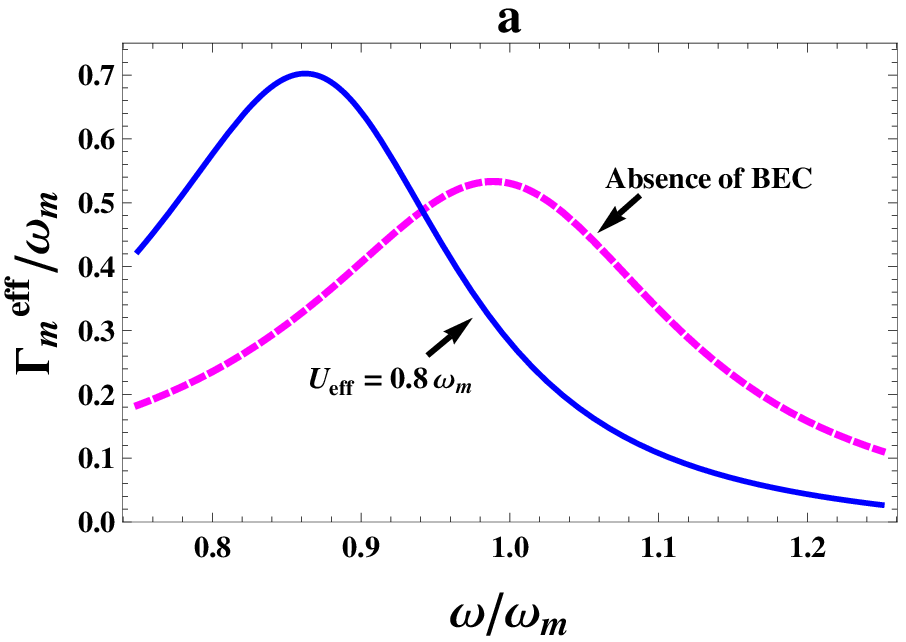}& \includegraphics [scale=0.80] {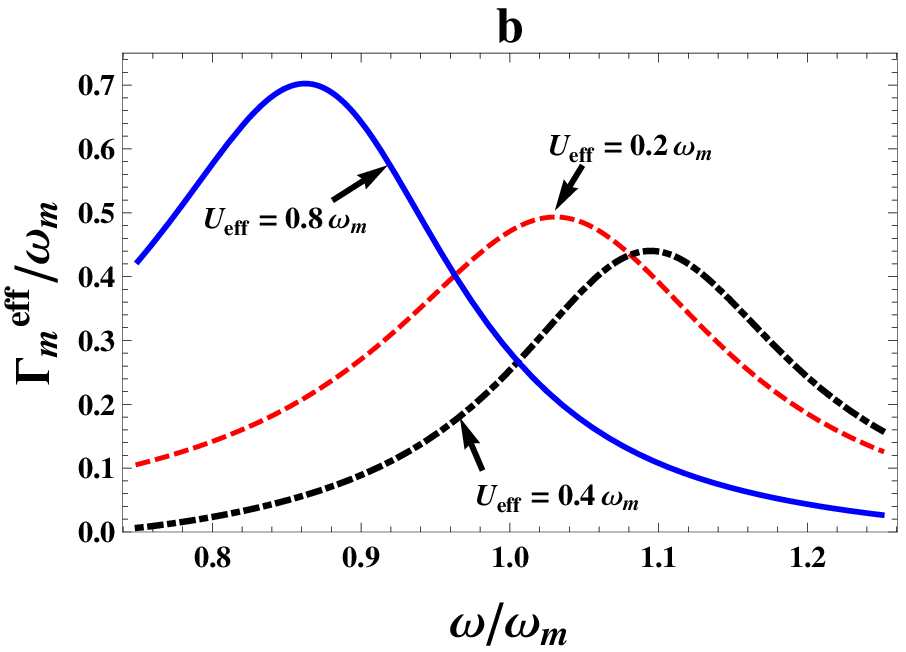}\\
\end{tabular}
\caption{(color online) Plot of normalized effective damping rate ($\Gamma_{m}^{eff}/\omega_{m}$) with dimensionless frequency in back-action cooling scheme. Fig. 3(a) represents the variation of dimensionless effective damping rate in the absence of BEC (dashed line) and in the presence of BEC (solid line) with $U_{eff}=0.8\omega_{m}$. Fig. 3(b) gives the deviation of dimensionless effective damping rate for three values of atomic two-body interaction with $U_{eff}=0.2\omega_{m}$ (dashed line), $U_{eff}=0.4\omega_{m}$ (dot dashed line) and $U_{eff}=0.8\omega_{m}$ (solid line). Other parameters chosen are the same as in figure 2.}
\label{3}
\end{figure}

\begin{figure}[h]
\hspace{-0.0cm}
\begin{tabular}{cc}
\includegraphics [scale=0.80]{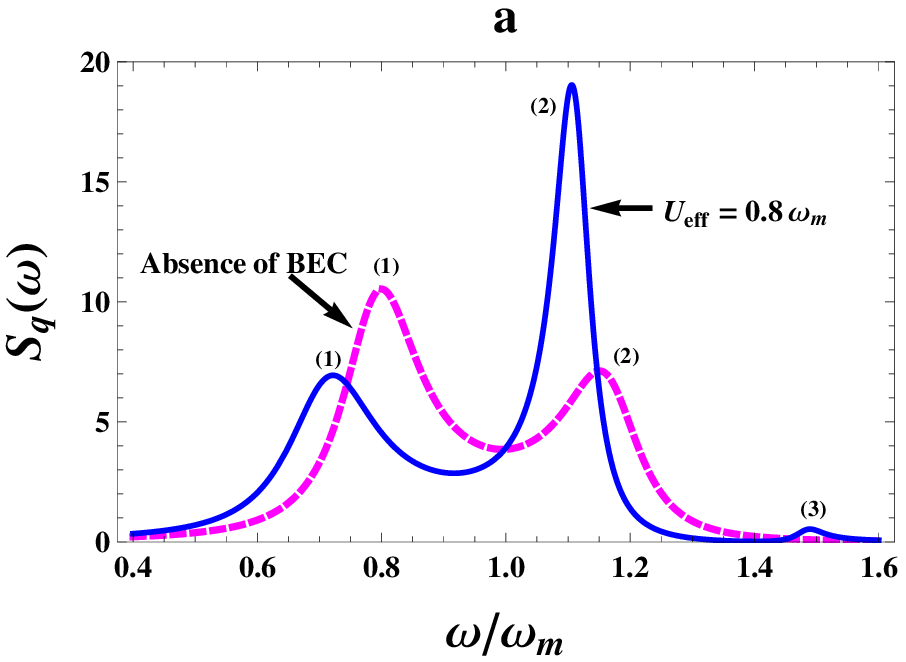}& \includegraphics [scale=0.80] {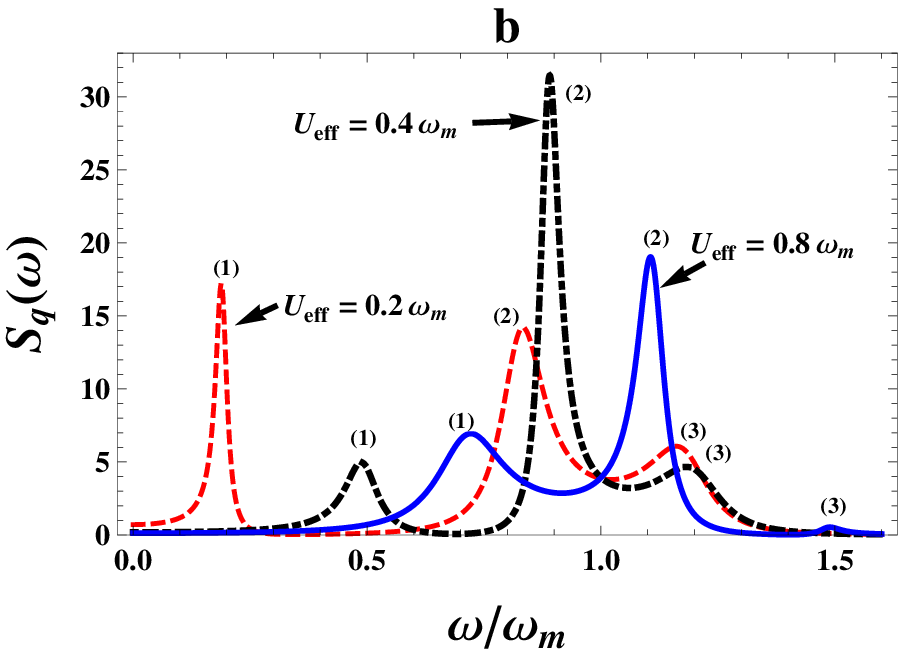}\\
\end{tabular}
\caption{(color online) Figure shows the displacement spectrum as a function of dimensionless frequency in back-action cooling scheme. Fig. 4(a) shows the displacement spectrum in the absence of BEC (dashed line) and in the presence of BEC (solid line) with $U_{eff}=0.8\omega_{m}$. Fig. 4(b) represents the displacement spectrum for three different values of atomic two-body interaction with $U_{eff}=0.2\omega_{m}$ (dashed line), $U_{eff}=0.4\omega_{m}$ (dot dashed line) and $U_{eff}=0.8\omega_{m}$ (solid line). $k_{B}T/\hbar\omega_{m}=10^{3}$ and the various other parameters used are same as in figure 2.}
\label{4}
\end{figure}

In Fig.2(a), the normalized effective mechanical frequency ($\omega_{m}^{eff}/\omega_{m}$) of the oscillating mirror in the absence of BEC (dashed line) is compared with the case in the presence of BEC (solid line) as a function of normalized frequency ($\omega/\omega_{m}$). This figure shows an extra resonance dip in the presence of BEC. Significant deviation from the bare frequency $\omega_{m}$ is observed around $\omega=\pm\omega_{m}$. This deviation is enhanced in the presence of BEC. Fig.2(b) illustrates the variation of normalized effective mechanical frequency of the mirror as a fuction of $\omega/\omega_{m}$ for different two body atom-atom interactions, $U_{eff}=0.2\omega_{m}$(dashed line), $U_{eff}=0.4\omega_{m}$(dot dashed line) and $U_{eff}=0.8\omega_{m}$(solid line). It can be seen from the figure that the deviation of the mechanical frequency of mirror from its resonance frequency $\omega_{m}$ decreases with the increase in condensate two-body interaction since the condensate becomes more robust with a higher two-body interaction. With in a set of experimentally achievable parameters, the atomic two-body interaction can be modified using the condensate cloud having dimensions $3.3$ $\mu$m \citep{30} ($290$ nm \citep{34}) with length $20$ $\mu$m \citep{30} ($615.5$ nm \citep{34}). It may also vary using scattering length ranging from $10 a_{0}$ to $190 a_{0}$ ($a_{0}$ = Bohr radius) \citep{57}. 

 Fig.3(a) shows the normalized effective mechanical damping ($\Gamma_{m}^{eff}/\omega_{m}$) of the mirror as a function of normalized  frequency ($\omega/\omega_{m}$) in the absence of BEC (dashed line) and the presence of BEC (solid line). As seen from this figure, below resonance ($\omega < \omega_{m}$), the effective damping of the oscillator is more in the presence of BEC whereas above resonance ($\omega > \omega_{m}$), it is less in the presence of BEC. Also, the plot of normalized effective mechanical damping of the mirror with normalized frequency is illustrated in the Fig.3(b) for the three different values of condensate two-body interactions, $U_{eff}=0.2\omega_{m}$(dashed line), $U_{eff}=0.4\omega_{m}$(dot dashed line) and $U_{eff}=0.8\omega_{m}$(solid line). A higher condensate two-body interaction enhances the effective damping of the mirror below resonance whereas it decreases the effective damping of the mirror above resonance. It shows an exception for the case $U_{eff}=0.4\omega_{m}$ as explained later. Fig.4(a) represents the displacement spectrum $S_{q}(\omega)$ as a function of dimensionless frequency ($\omega/\omega_{m}$) in the presence of BEC (solid line) and the absence of BEC (dashed line). We observe the normal mode splitting into two modes in the absence of BEC and we find that the normal mode splits up into three modes in the presence of BEC. This extra mode is the result of additional quantum fluctuations of the condensate (Bogoliubov mode). Also below resonance, we observe that the amplitude of the peak(1) of displacement spectrum in the absence of BEC is greater than that in the presence of BEC. This is due to the higher damping rate in the presence of BEC as illustrated in Fig.3(a). This conclusion is exactly opposite to that for the case above resonance. Fig.4(a) depicts that the amplitude of peak (1) in the absence of BEC is more than with BEC whereas the amplitude of peak (2) in the absence of BEC is less than with BEC. This represents the energy exchange between different modes of the system. Also, Fig.4(b) illustrates the displacement spectrum varying with normalized frequency for three different values of condensate two body interaction, $U_{eff}=0.2\omega_{m}$(dashed line), $U_{eff}=0.4\omega_{m}$(dot dashed line) and $U_{eff}=0.8\omega_{m}$(solid line). Here coupling between the mechanical mode of the mirror, fluctuation of cavity field around steady state and the fluctuations of the condensate (Bogoliubov mode) around the mean field results in normal-mode splitting (NMS). For the observation of NMS, it is significant to note that the time scale for the exchange of energy between the mechanical mode, photon mode and the Bogoliubov mode should be faster than the decoherence of each mode. The beat of laser pump photons with the photons scattered from the condensate atoms is the source of cavity field fluctuations. Since the frequency of the Bogoliubov mode of the condensate is directly proportional to $\sqrt{U_{eff}}$, fig.4(b) shows a variation in the displacement spectrum for different $U_{eff}$. The spectrum shifts towards the right with the increase in condensate two body interaction. Recently, an experiment reveals that Bogoliubov mode of cloud of ultracold atoms interacting with optical resonator has momentum $\pm 2 k_{c}$($k_{c}$ is the cavity wave number) \citep{30}.
 
The condition $\Gamma_{m}<<\omega_{m}<<k_{B}T/\hbar$ is always considered in standard optomechanical experiments \citep{22,51,60,61}. In this limiting case, we consider the approximation that $\coth \left (\hbar \omega/2 k_{B}T\right) \simeq 2k_{B}T/\hbar\omega$. Now, in order to acheive the ground state cooling of the mechanical resonator, we measure the average energy of the resonator in steady state, which is given by \citep{50}

\begin{equation}\label{energy}
U=\frac{\hbar \omega_{m}}{2} \left[\langle \delta q^{2} \rangle 
+ \langle \delta p^{2} \rangle \right] =\hbar \omega_{m} \left(n_{eff}+\frac{1}{2} \right).
\end{equation}

\begin{figure}[h]
\hspace{-0.0cm}
\begin{tabular}{cc}
\includegraphics [scale=0.80]{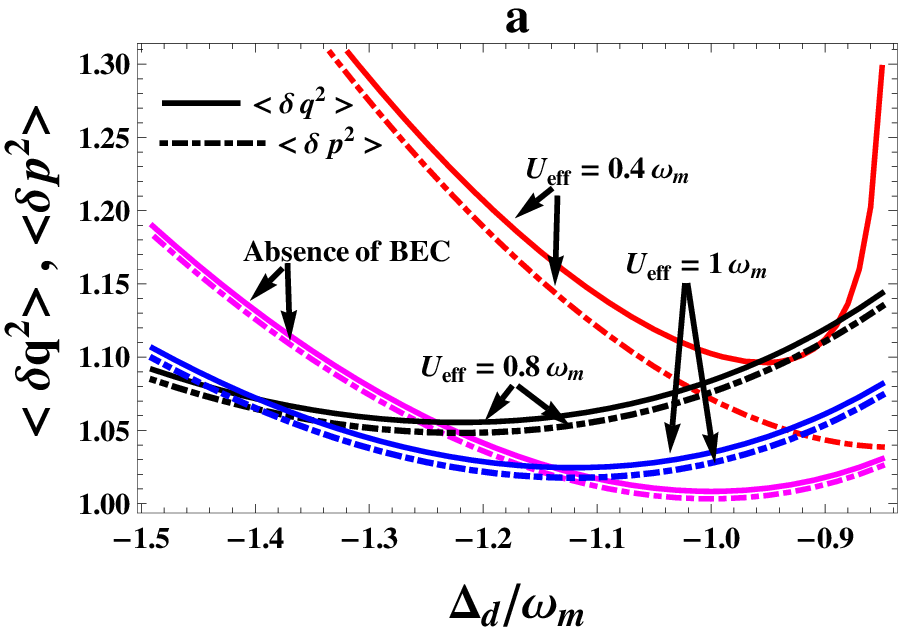}& \includegraphics [scale=0.80] {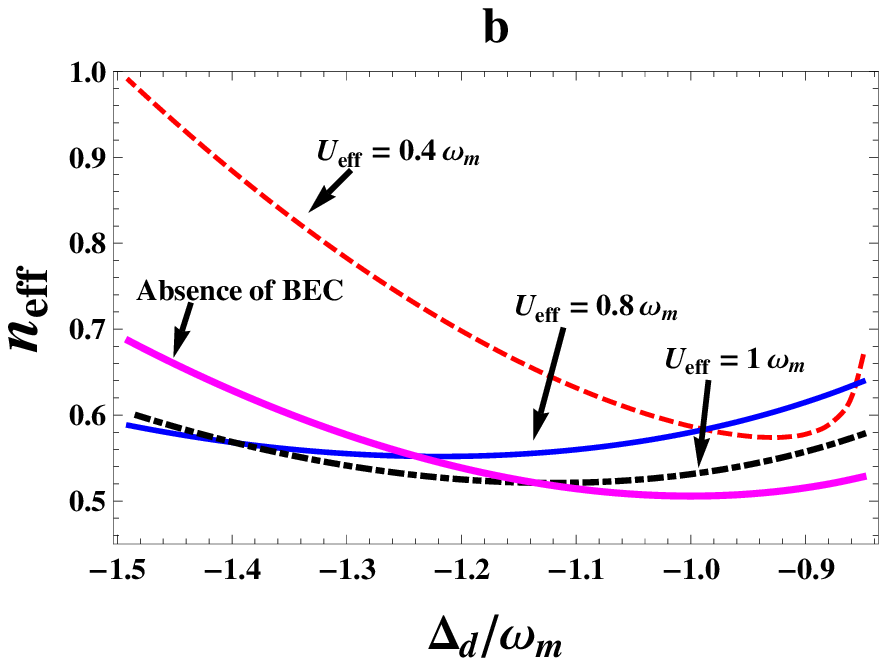}\\
\end{tabular}
\caption{(color online) Back-action cooling scheme: (a) Plot of the mirror's position variances $\left\langle \delta q^{2}\right\rangle $ (Solid line) and the mirror's momentum variances $\left\langle \delta p^{2}\right\rangle $ (Dot dashed line) versus the normalized effective detuning ($\Delta_{d}/\omega_{m}$) in the absence of BEC and for three different values of atomic two-body interaction with $U_{eff}=0.4\omega_{m}$, $U_{eff}=0.8\omega_{m}$ and $U_{eff}=\omega_{m}$. Other parameters chosen are: $\Gamma_{m}=10^{-7}\omega_{m}$, $\kappa=0.1\omega_{m}$, $G=\omega_{m}$, $\beta=0.05$, $\nu=0.01\omega_{m}$, $g_{c}=0.5\omega_{m}$ and $k_{B}T/\hbar\omega_{m}=10^{3}$. (b) Effective phonon number with dimensionless effective detuning in the absence of BEC (thick Solid line) and for three different values of atomic two-body interaction with $U_{eff}=0.4\omega_{m}$ (dashed line), $U_{eff}=0.8\omega_{m}$ (thin Solid line) and $U_{eff}=\omega_{m}$ (dot dashed line). Other parameters chosen are the same as in (a).}
\label{5}
\end{figure}

\begin{figure}[h]
\hspace{-0.0cm}
\includegraphics [scale=0.8]{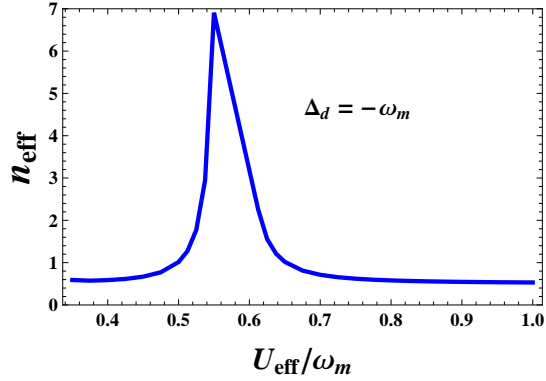}
\caption{(color online) The effective phonon number versus the normalized atomic two-body interaction in back-action cooling scheme at $\Delta_{d}=-\omega_{m}$. Rest of the parameters are same as in figure 5.}
\label{6}
\end{figure}

\begin{figure}[h]
\hspace{-0.0cm}
\includegraphics [scale=0.8]{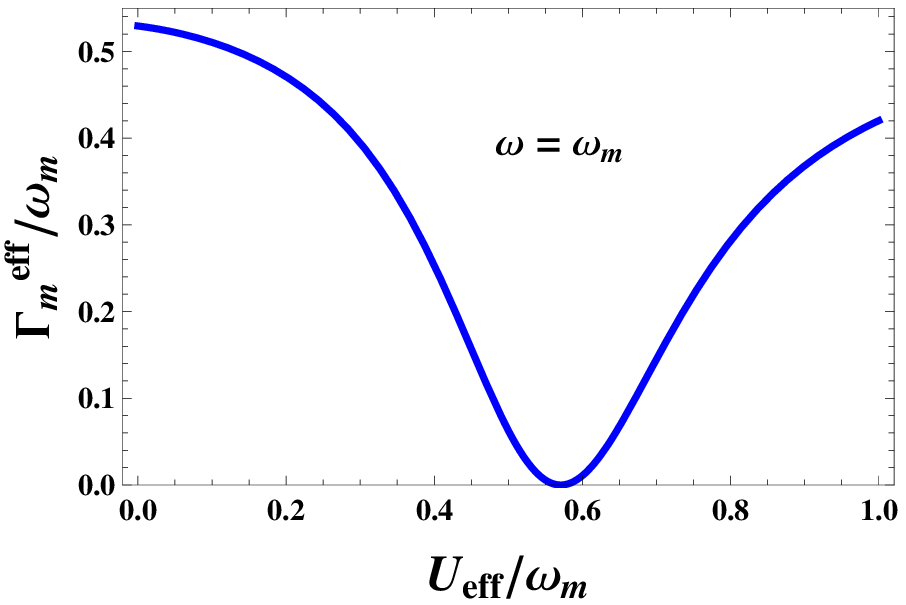}
\caption{(color online)The normalized effective mechanical damping rate versus the normalized atomic two-body interaction at the resonance frequency ($\omega=\omega_{m}$) in back-action cooling scheme. Other parameters chosen are same as figure 2.}
\label{7}
\end{figure}

The system has to be stable in order to be in steady state. Hence, the stability conditions for back-action cooling scheme given in the Appendix II must always be satisfied in this regime such that all the poles of $\chi_{eff}(\omega)$ lie in the lower complex half-plane. Here, equation  (\ref{energy}) implies that the effective phonon number of the mirror's motion can be written as

\begin{equation}\label{neff}
n_{eff}=\frac{1}{2} \left[\langle \delta q^{2} \rangle 
+ \langle \delta p^{2} \rangle - 1 \right],
\end{equation} 

where $\langle \delta q^{2} \rangle$ and $\langle \delta p^{2} \rangle$ represent the displacement and momentum variances of the oscillator respectively which are defined as \citep{12}

\begin{equation}\label{delq}
\langle \delta q^{2} \rangle =\int_{-\infty}^{\infty} \frac{d\omega}{2 \pi} S_{q}(\omega),
\end{equation} 

\begin{equation}\label{delp}
\langle \delta p^{2} \rangle =\int_{-\infty}^{\infty} \frac{d\omega}{2 \pi} \frac{\omega^{2}}{\omega_{m}^{2}} S_{q}(\omega).
\end{equation}

We have solved these oscillator variances numerically using the MATHEMATICA 8.0. Generally, there is no equipartition of energy, $\langle \delta q^{2} \rangle \neq \langle \delta p^{2} \rangle$. In order to approach the ground state cooling, the effective phonon number should be less than one i.e., $n_{eff} < 1$ which can only be attained if the initial average-thermal excitation number $\overline{n}=\left[ exp\left\lbrace \frac{\hbar\omega_{m}}{k_{B}T}\right\rbrace -1 \right]^{-1}$ is not excessively large. This can be possible even at cryogenic temperatures if $\omega_{m}$ is sufficiently large.

In order to get an intuitive picture, we have plotted both the displacement and momentum variances of the mirror as a function of dimensionless  detuning $\Delta_{d}/\omega_{m}$ in the absence and presence of BEC as illustrated in Fig.5(a). Fig.5(a) shows that, for large values of detuning, the variances have lesser values in the absence of BEC than in the presence of BEC. It is evident from the figure that one can access the gound state of the system by increasing the Bose-Einstein condensate two-body interaction. This implies that $U_{eff}$ may alter the cooling process significantly. So, in order to acheive the best condition for the ground state cooling of the system, $U_{eff}$ can be used as an additional control parameter. Fig.5(b) depicts the effective phonon number $n_{eff}$ as a function of dimensionless detuning $\Delta_{d}/\omega_{m}$ for the absence of BEC (thick solid line) and the three different values of $U_{eff}$, $U_{eff}=0.4\omega_{m}$(dashed line), $U_{eff}=0.8\omega_{m}$(thin solid line) and $U_{eff}=1\omega_{m}$(dot dashed line). Clearly, with the increase in $U_{eff}$, the minimum value of $n_{eff}$ is shifting towards zero such that, for $U_{eff}=1 \omega_{m}$, we have $n_{eff} \simeq 0.52$ at $\Delta_{d}=-1.13\omega_{m}$. However, the least value of $n_{eff}$ is reached in the absence of BEC which is nearly $0.51$ at $\Delta_{d}= -1 \omega_{m}$. Hence, in back-action cooling scheme, we have noticed that the minimum value of effective phonon number is found in the absence of BEC. Moreover, we have observed that for small values of $\Delta_{d}$, presence of BEC gives better results for higher $U_{eff} (= 0.8 \omega_{m}$  and  $\omega_{m})$ than the absence of BEC. One can also observe from Fig.5(b), the effective temperature of the oscillator does not vary much in the presence of BEC($U_{eff}=0.8\omega_{m}$ and $\omega_{m}$) as compared to that in the absence of BEC. This reveals that for a wide range of detuning($\Delta_{d}$), the effective temperature of the mirror is not changing significantly for higher $U_{eff}$. Therefore, one can use condensate two-body interaction as a new handle to acheive and sustain a low temperature of the mirror with BEC over a wide range of $\Delta_{d}$. The atomic two-body interaction is proportional to number of atoms ($N$). In the recent past, \citep{56} has shown experimentally that increasing the number of atoms enhances the damping of the oscillating membrane which is coupled to BEC through cavity field. This validates our work that the energy of the mirror decreases by increasing $U_{eff}$. Further, it is noticed that both the variances tend to $\left\langle \delta q^{2} \right\rangle \simeq \left\langle \delta p^{2} \right\rangle \simeq 1/2$ for $G=0.5\omega_{m}$ (keeping other parameters same) i.e., energy equipartiton is satisfied in this chosen parameter regime but it results in the approximate same ground state oscillator energy as we have acheived in the above mentioned general case of no energy equipartition ($\left\langle \delta q^{2} \right\rangle \neq \left\langle \delta p^{2} \right\rangle$). We find that the best cooling regime is acheived in the good cavity limit condition ($\kappa << \omega_{m}$) by taking $\kappa=0.1 \omega_{m}$ as mentioned in \citep{12}.

Variation of effective phonon number with $U_{eff}/\omega_{m}$ at $\Delta_{d}= -\omega_{m}$ is also shown in Fig.6. It depicts a rapid increase in $n_{eff}$ for a range of $U_{eff}$. This corresponds to the sudden decrease in the effective damping rate of mirror as illustrated in Fig.7. It is by virtue of the fact that, in the range $0\leq U_{eff} \leq 0.57 \omega_{m}$, the second term in Eqn. (\ref{gamma}) increases as ($\omega^{2}-\beta_{1}\beta_{2}$) decreases. On the other hand, for  $U_{eff}>0.57 \omega_{m}$, this term decreases since ($\omega^{2}-\beta_{1}\beta_{2}$) becomes negative. It explains the exception observed in Fig. 3(b) and 4(b) for $U_{eff}=0.4\omega_{m}$. In the next section, we study the system using the cold damping feedback technique.

\section{Cold Damping Feedback Scheme}

Cold damping feedback technique is an alternative method to improve the cooling of a mechanical oscillator by overdamping it at the quantum level without increasing the thermal noise of the system as proposed in \citep{16,26,27}. This technique has been realized experimentally \citep{22,23,51}. It involves negative derivative feedback technique. The displacement of the oscillator is measured through phase-sensitive homodyne detection of the cavity output which is fed back to the resonator using a force proportional to the oscillator velocity \citep{10,12}.

\begin{figure}[h]
\hspace{-0.0cm}
\includegraphics [scale=0.8]{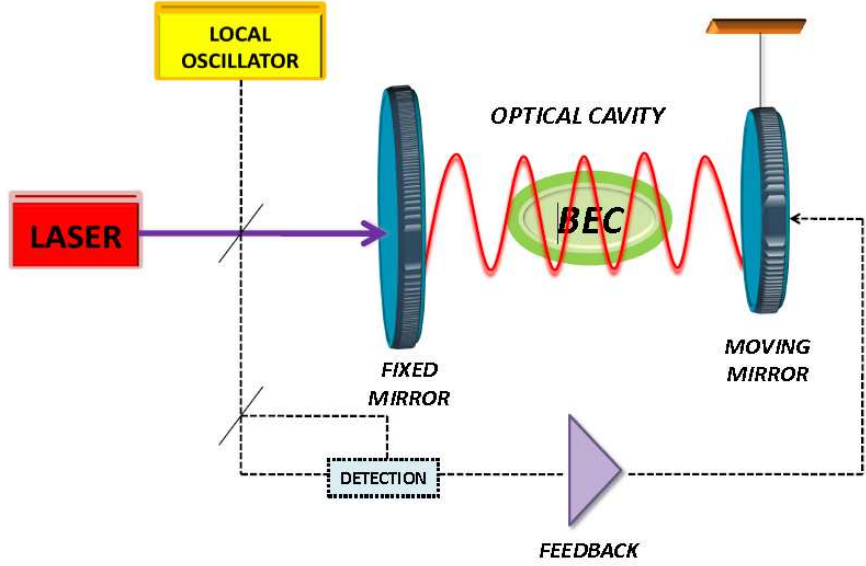}
\caption{(color online) Setup for Cold Damping. The configuration is same as in case of Back-Action Cooling involving an additional feedback loop with a force proportional to oscillator velocity. Cavity output field is homodyne detected using the beam splitter. }
\label{8}
\end{figure}

Therefore, the QLEs involving a feedback force for this scheme are given by :

\begin{equation}
\delta \dot{q_{a}}(t)=\beta_{1} \delta p_{a}(t),
\end{equation}

\begin{equation}
\delta \dot{p_{a}}(t)=-\beta_{2} \delta q_{a}(t)-2 g_{c}\delta q_{b}(t),
\end{equation}

\begin{equation}\label{X}
\delta \dot{q_{b}}(t)=-\frac{\kappa}{2} \delta q_{b}(t)+\sqrt{\kappa} q_{in}(t)-\Delta_{d} \delta p_{b}(t),
\end{equation}

\begin{equation}
\delta \dot{p_{b}}(t)=-\frac{\kappa}{2} \delta p_{b}(t)-2 g_{c} \delta q_{a}(t)+2G \beta \delta q(t)+\sqrt{\kappa}p_{in}(t)+\Delta_{d} \delta q_{b}(t),
\end{equation}

\begin{equation}
\delta \dot{q}(t)=\omega_{m} \delta p(t),
\end{equation}

\begin{equation}\label{p}
\delta \dot{p}(t)=-\omega_{m} \delta q(t)+ 2 G \beta \delta q_{b}(t)-\Gamma_{m} \delta p(t)+W(t)- \int\nolimits_{-\infty}^{t} ds g(t-s) \delta p_{est}(s).
\end{equation}

where $\beta_{1}=\nu+U_{eff}$ , $\beta_{2}=\nu+ 3U_{eff}$ and the filter function $g(t)$ is the causal kernel chosen as \citep{12}
\begin{equation}
g(t)=g_{cd} \frac{d}{dt} [\theta(t)\omega_{fb} e^{-\omega_{fb}t}],
\end{equation}
such that
\begin{equation}\label{g}
g(\omega)=\frac{-i\omega g_{cd} \sqrt{\lambda}}{1-i\omega/\omega_{fb}},
\end{equation}
where $g(\omega)$ is the Fourier transform of $g(t)$. Here $g_{cd}$ is the positive feedback gain and $\lambda$ quantifies the homodyne detection efficiency of the photodetector. Generally the detector efficiency is $\lambda<1$ for the inefficient homodyne detection if the effect of additional noise is considered, but for ideal homodyne detection, $\lambda=1$ \citep{52}. $\omega_{fb}^{-1}$ signifies the feedback loop delay time. Moreover, $\delta p_{est}(s)$ represents the estimated intracavity phase quadrature which is evaluated by using the generalized input-output phase relation given by \citep{52,53,54} 
\begin{equation}
p_{out}(t)= \sqrt{\kappa} \sqrt{\lambda} \delta p_{b}(t)- \sqrt{\lambda} p_{in}(t) - \sqrt{1-\lambda} p_{v}(t).
\end{equation} 

Here $p_{out}(t)$ and $p_{in}(t)$ represent the output homodyne phase quadrature and input noise respectively. Also $p_{v}(t)$ is the associated vaccum field quadrature which describes the additional noise for $\lambda<1$ such that 
\begin{equation}
\delta p_{est}(t)=\frac{p_{out}(t)}{\sqrt{\kappa}}= \sqrt{\lambda} \delta p_{b}(t)-\frac{\sqrt{\lambda} p_{in}(t)}{\sqrt{\kappa}}  -\frac{ \sqrt{1-\lambda} p_{v}(t)}{\sqrt{\kappa}},
\end{equation}

where $p_{v}(t)$ and $p_{in}(t)$ are uncorrelated. Now the QLEs supplemented with the feedback term are solved in the frequency domain such that the displacement and momentum variances of the oscillator are given by the Eqns.(\ref{delq}) and (\ref{delp}) respectively. It gives a distinct displacement spectrum using the correlations given in Appendix I. Explicitly, it can be written as

\begin{equation}
S_{q}^{cd}(\omega)=\vert \chi_{eff}^{cd}(\omega)\vert^{2} [S_{th}(\omega) + S_{rp}(\omega , \Delta_{d}) + S_{fb}(\omega)].
\end{equation} 

Here, $S_{th}(\omega)$ and $S_{rp}(\omega , \Delta_{d})$ are given by the Eqns. (\ref{th}) and (\ref{rp}) respectively. In this scheme, the position spectrum consists of an additional feedback-induced term expressed as

\begin{equation}
\begin{split}
S_{fb}(\omega)=\frac{\left[2G\beta \sqrt{\lambda}(\omega^{2}-\beta_{1}\beta_{2})\left\lbrace \Delta_{d}(\omega^{2}-\beta_{1}\beta_{2})(\omega^{2}-\frac{\kappa^{2}}{4}-\Delta_{d}^{2})+4g_{c}^{2}\beta_{1}(\Delta_{d}^{2}+\frac{\kappa^{2}}{2}) \right\rbrace \left\lbrace g(\omega)+g(-\omega)\right\rbrace \right] }{X(\omega)}\\-\frac{\left[ 2iG \omega \beta \kappa \sqrt{\lambda}(\omega^{2}-\beta_{1}\beta_{2})\left\lbrace  \Delta_{d}(\omega^{2}-\beta_{1}\beta_{2})-4g_{c}^{2}\beta_{1}\right\rbrace \left\lbrace g(\omega)-g(-\omega)\right\rbrace\right] }{X(\omega)}\\+\vert g(\omega)\vert^{2}\left\lbrace \frac{1}{\kappa}-\frac{\left[ 4g_{c}^{2}\beta_{1}\lambda\kappa\left\lbrace \Delta_{d}(\omega^{2}-\beta_{1}\beta_{2})-4g_{c}^{2}\beta_{1}\right\rbrace\right] }{X(\omega)}\right\rbrace,
\end{split}
\end{equation}

which arises since the cold damping loop feeds back the measurement noise into the dynamics of the movable mirror. $\chi_{eff}^{cd}(\omega)$ is the effective mechanical susceptibility of the movable mirror, modified by the filter function, given by
 
\begin{equation}
\chi_{eff}^{cd}(\omega)=\frac{\omega_{m}}{\left[(\omega_{m}^{2}-\omega^{2}+i \omega \Gamma_{m})+\chi_{1}^{cd}(\omega)\right]},
\end{equation}

where
\begin{equation}
\chi_{1}^{cd}(\omega)=\frac{2G\beta\omega_{m}(\omega^{2}-\beta_{1}\beta_{2})\left\lbrace g(-\omega)\sqrt{\lambda}(i\omega +\frac{\kappa}{2})+2G\beta \Delta_{d}\right\rbrace}{\left\lbrace (\omega^{2}-\beta_{1}\beta_{2})\left( \Delta_{d}^{2}+\frac{\kappa^{2}}{4}-\omega^{2}+i\omega \kappa\right)-4g_{c}^{2}\Delta_{d}\beta_{1}\right\rbrace}.
\end{equation}

It gives us the effective resonance frequency and damping rate using Eqn. (\ref{g}):

\begin{eqnarray}\label{wcd}\nonumber
\omega_{m}^{eff,cd}(\omega)=\left[ \omega_{m}^{2}+\omega_{m}^{op,cd}\right]^{1/2},
\end{eqnarray}

where

\begin{eqnarray}\label{opcd}\nonumber
\omega_{m}^{op,cd} &=& X_{1}(\omega)\left[ (\omega^{2}-\beta_{1} \beta_{2})[\Delta_{d}^{2}+\frac{\kappa^{2}}{4}-\omega^{2}] -4 g_{c}^{2} \Delta_{d} \beta_{1} \right]\left( 4G\beta \Delta_{d}+\frac{2 \omega^{2} g_{cd} \omega_{fb}\lambda}{(\omega^{2}+\omega_{fb}^{2})}(\frac{\kappa}{2}-\omega_{fb})\right)\nonumber \\ &+& X_{1}(\omega)(\omega^{2}-\beta_{1} \beta_{2})\left(\frac{2 \omega^{2} g_{cd} \omega_{fb}\lambda \kappa}{(\omega^{2}+\omega_{fb}^{2})}(\omega^{2}+\frac{\omega_{fb}\kappa}{2}) \right),
\end{eqnarray}

\begin{eqnarray}\label{gcd}\nonumber
\Gamma_{m}^{eff,cd}(\omega) &=& \Gamma_{m}+ X_{1}(\omega)\left[ (\omega^{2}-\beta_{1} \beta_{2})[\Delta_{d}^{2}+\frac{\kappa^{2}}{4}-\omega^{2}]-4 g_{c}^{2} \Delta_{d} \beta_{1} \right]\left( \frac{2 g_{cd} \omega_{fb} \lambda}{(\omega^{2}+\omega_{fb}^{2})}(\omega^{2}+\frac{\omega_{fb}\kappa}{2})\right)\nonumber \\ &-& X_{1}(\omega)\kappa (\omega^{2}-\beta_{1} \beta_{2})\left(4G\beta \Delta_{d}+\frac{2 \omega^{2} g_{cd} \omega_{fb}\lambda}{(\omega^{2}+\omega_{fb}^{2})}(\frac{\kappa}{2}-\omega_{fb})\right),
\end{eqnarray}

where
\begin{equation}
X_{1}(\omega)=\frac{G \beta \omega_{m} (\omega^{2}-\beta_{1} \beta_{2})}{X(\omega)}.
\end{equation}

\begin{figure}[h]
\hspace{-0.0cm}
\begin{tabular}{cc}
\includegraphics [scale=0.80]{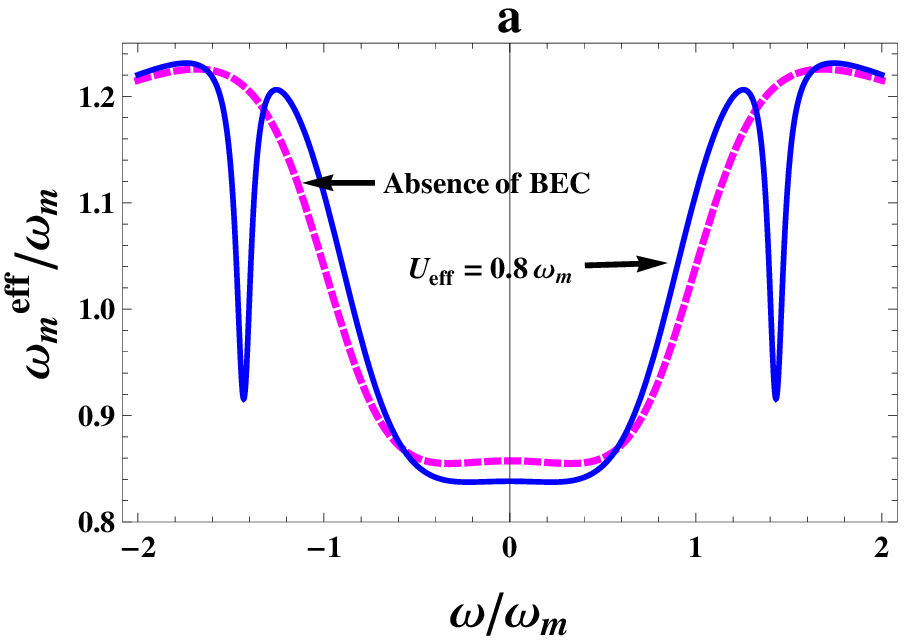}& \includegraphics [scale=0.80] {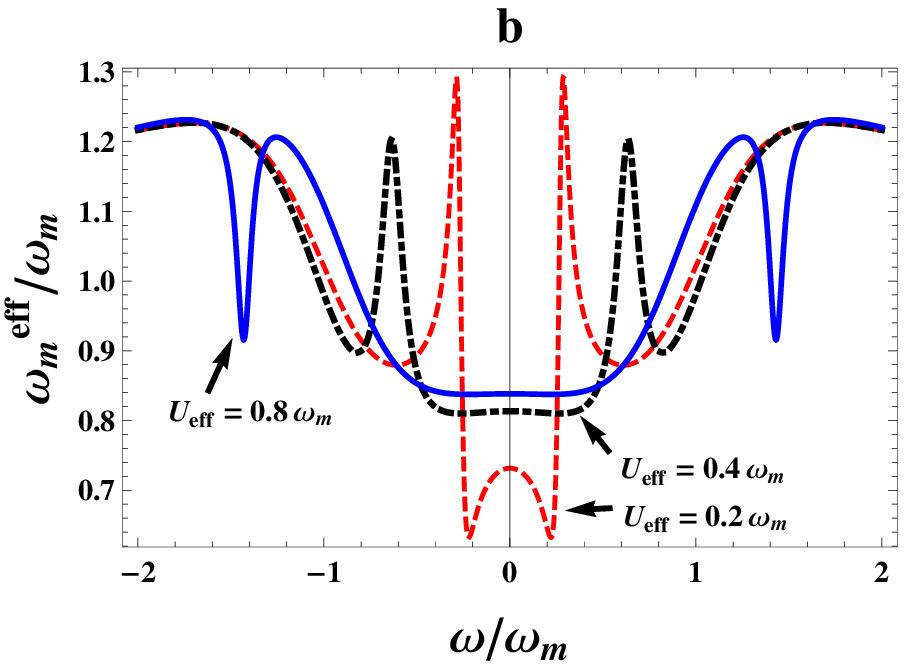}\\
\end{tabular}
\caption{(color online) Plot of normalized effective mechanical frequency ($\omega_{m}^{eff}/\omega_{m}$) as a function of dimensionless frequency in cold damping feedback scheme. Fig. 9(a) shows the variation of the dimensionless effective mechanical frequency in the absence of BEC (dashed line) and in the presence of BEC (solid line) with $U_{eff}=0.8\omega_{m}$. Fig. 9(b) represents the dimensionless effective mechanical frequency for three different values of atomic two-body interaction with $U_{eff}=0.2\omega_{m}$ (dashed line), $U_{eff}=0.4\omega_{m}$ (dot dashed line) and $U_{eff}=0.8\omega_{m}$ (solid line). Other parameters used are : $\Gamma_{m}=10^{-5}\omega_{m}$, $\Delta_{d}=-\omega_{m}$, $\kappa=1.2\omega_{m}$, $G=6\omega_{m}$, $\beta=0.05$, $\nu=0.015\omega_{m}$, $g_{c}=0.3\omega_{m}$, $\omega_{fb}=4\omega_{m}$, $g_{cd}=0.8$, $\lambda=0.8$ and $k_{B}T/\hbar\omega_{m}=10^{3}$.}
\label{9}
\end{figure}

\begin{figure}[h]
\hspace{-0.0cm}
\begin{tabular}{cc}
\includegraphics [scale=0.80]{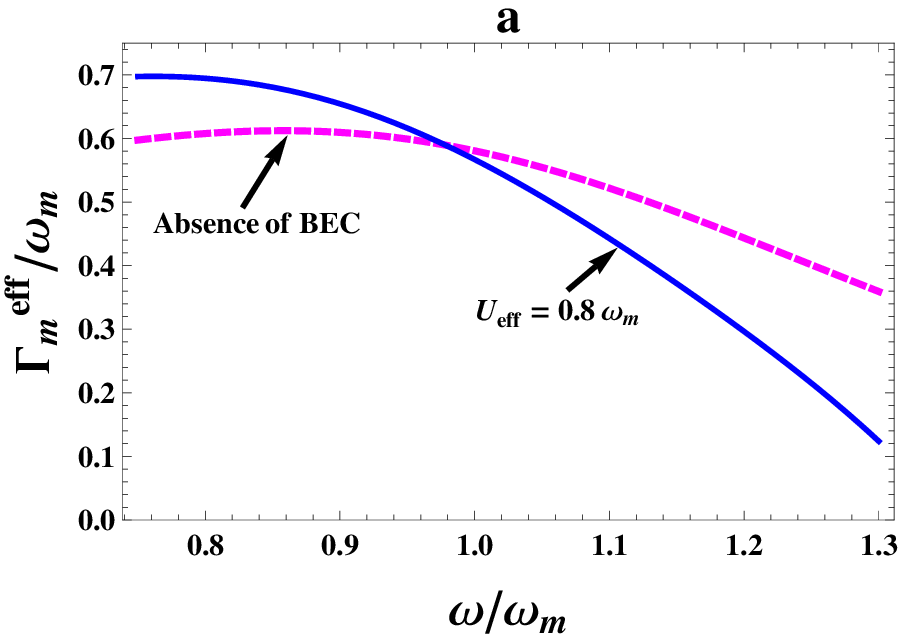}& \includegraphics [scale=0.80] {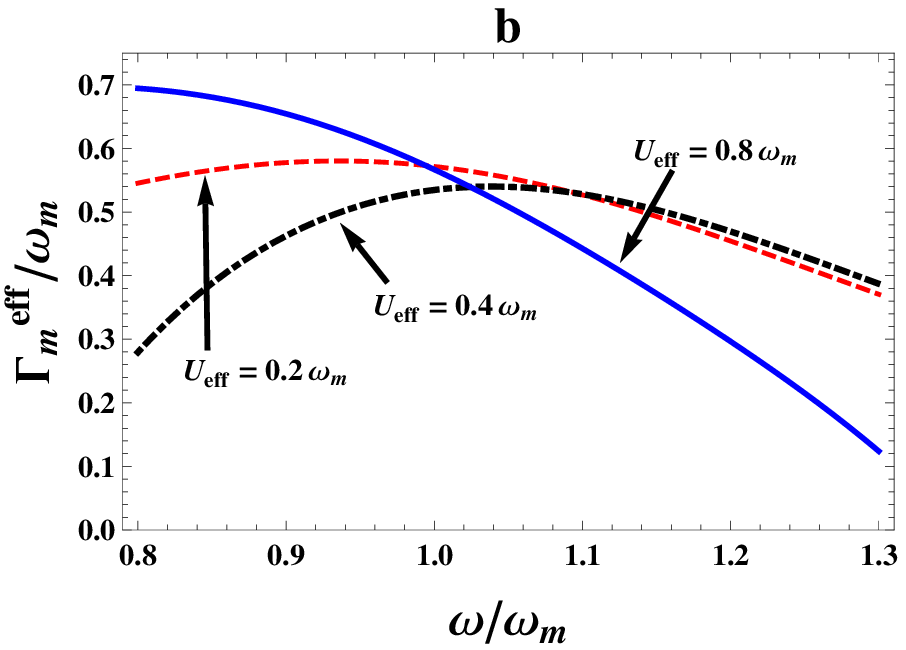}\\
\end{tabular}
\caption{(color online) Plot of normalized effective damping rate ($\Gamma_{m}^{eff}/\omega_{m}$) with dimensionless frequency in cold damping feedback scheme. Fig. 10(a) represents the variation of dimensionless effective mechanical frequency in the absence of BEC (dashed line) and in the presence of BEC (solid line) with $U_{eff}=0.8\omega_{m}$. Fig 10(b) gives the deviation of dimensionless effective mechanical frequency for three values of atomic two-body interaction with $U_{eff}=0.2\omega_{m}$ (dashed line), $U_{eff}=0.4\omega_{m}$ (dot dashed line) and $U_{eff}=0.8\omega_{m}$ (solid line). Other parameters chosen are same as in figure 9.}
\label{10}
\end{figure}

\begin{figure}[h]
\hspace{-0.0cm}
\includegraphics [scale=0.8]{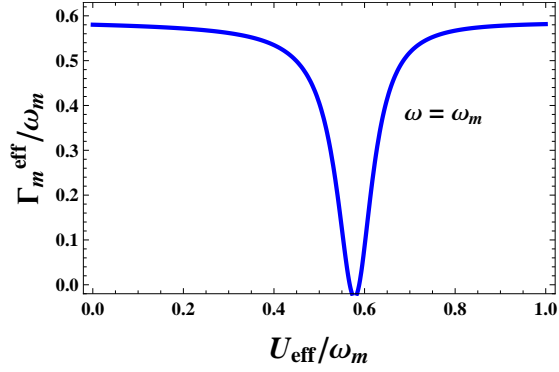}
\caption{(color online) The normalized effective mechanical damping rate versus the normalized atomic two-body interaction at the resonance frequency ($\omega=\omega_{m}$) in cold damping feedback scheme. Other parameters chosen are same as figure 9.}
\label{11}
\end{figure}

\begin{figure}[h]
\hspace{-0.0cm}
\begin{tabular}{cc}
\includegraphics [scale=0.80]{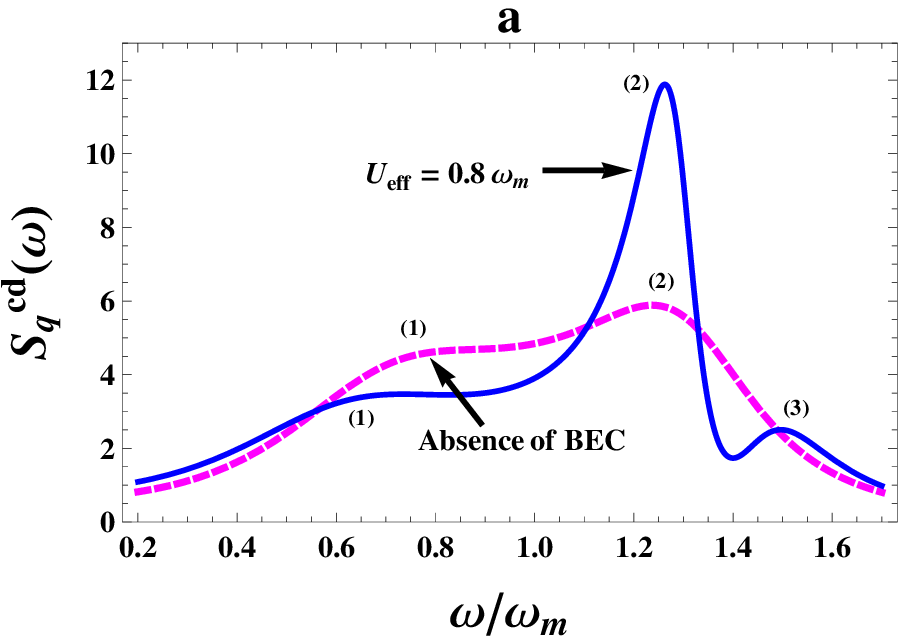}& \includegraphics [scale=0.80] {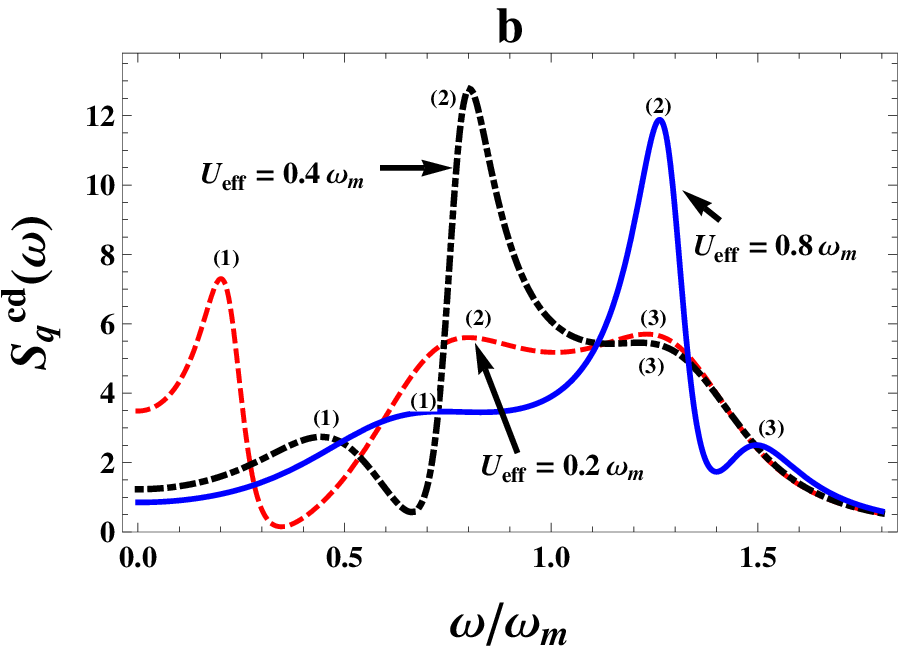}\\
\end{tabular}
\caption{(color online) Figure shows the displacement spectrum as a function of dimensionless frequency in cold damping feedback scheme. Fig. 11(a) shows the displacement spectrum in the absence of BEC (dashed line) and in the presence of BEC (solid line) with $U_{eff}=0.8\omega_{m}$. Fig. 11(b) represents the displacement spectrum for three different values of atomic two-body interaction with $U_{eff}=0.2\omega_{m}$ (dashed line), $U_{eff}=0.4\omega_{m}$ (dot dashed line) and $U_{eff}=0.8\omega_{m}$ (solid line). $k_{B}T/\hbar\omega_{m}=10^{3}$ and the various other parameters used are same as figure 9.}
\label{12}
\end{figure}

In order to overdamp the mechanical mirror oscillations, an additional viscous force known as the feedback force is applied in this technique which is possible only when the estimated intracavity phase quadrature $\delta p_{est}$ is proportional to the oscillator position $\delta q(t)$. It can be achieved in the bad cavity limit where $\kappa>\omega_{m}$. Hence, for cold damping feedback, we limit our discussion to the bad cavity limit. We have plotted the effective resonance frequency ($\omega_{m}^{eff,cd}$) and damping rate ($\Gamma_{m}^{eff,cd}$) as a function of dimensionless frequency ($\omega/\omega_{m}$) in Figs.(9) and (10) respectively in order to compare them with the back-action cooling plots. It can be seen again from the fig.(9) that there is no significant shift in the frequency in both absence and presence of BEC in the chosen parameter regime. Hence, there is negligible optical spring effect as $\omega_{m}^{2}$ dominates over $\omega_{m}^{op,cd}$ for higher resonance frequencies. However, the mechanical damping rate shows significant variation with change in frequency. Below resonance, effective damping increases by adding BEC to the system [see Fig.10(a)] and by increasing the atom-atom interaction [see Fig.10(b)] with an exception for $U_{eff}=0.4\omega_{m}$. This exception in the result can be explained from the Fig.(11) which clearly shows the variation in mechanical damping rate with increasing $U_{eff}$. The reason for this sudden decrease in the damping rate in a particular region of $U_{eff}$ for $\omega=\omega_{m}$ is similar as we have described in case of back-action cooling. However, in the zero detuning case, one can clearly observe from the expressions ($\ref{wcd}$) and ($\ref{gcd}$) that $\omega_{m}^{eff,cd}$ and $\Gamma_{m}^{eff,cd}$ are independent of BEC parameters and behave in the same manner as in the case for absence of BEC. In the adiabatic limit for zero detuning i.e., $\kappa,\omega_{fb}>>\omega$, we get $\omega_{m}^{eff,cd}\simeq \omega_{m}$ and $\Gamma_{m}^{eff,cd}=\Gamma_{m}+\frac{g_{cd} G \beta \omega_{m}\lambda}{4 \kappa}$. This shows that the effective damping rate increases in this scheme without involving any significant change in the resonance frequency of the oscillator. We have also shown the plot of displacement spectrum ($S_{q}^{cd}(\omega)$) as a function of dimensionless frequency ($\omega/\omega_{m}$) in Fig.12 to compare it with the corresponding curve for the back-action cooling. This scheme involves an additional feedback-induced term denoted by $S_{fb}(\omega)$ which gives a distinct displacement spectrum than the back-action cooling scheme. This feedback induced term $S_{fb}(\omega)$ can be used as an additional handle in the cold damping feedback scheme. The displacement spectrum in this scheme can be manipulated by a coherent control over the feedback parameters $g_{cd}$ and $\omega_{fb}$. The variation in amplitude of the peaks in the absence and presence of BEC of the displacement spectrum, shown in fig.12(a), represents the energy exchange between the different modes of the system. Only two-mode splitting is observed in the absence of BEC [see Fig.12(a)] while normal mode splits up into three modes due to atom-atom interaction [see Fig.12(b)].

\begin{figure}[h]
\hspace{-0.0cm}
\begin{tabular}{cc}
\includegraphics [scale=0.80]{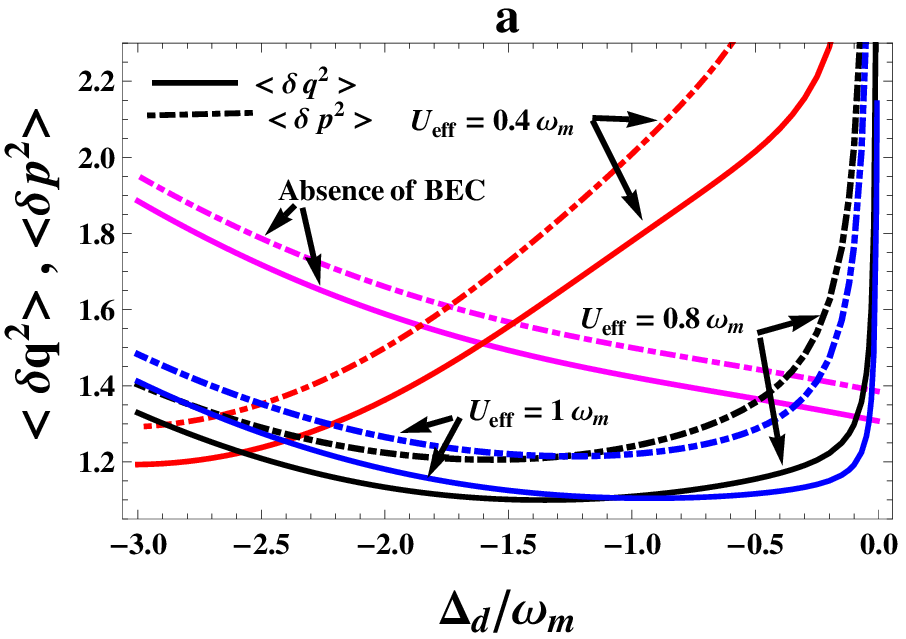}& \includegraphics [scale=0.80] {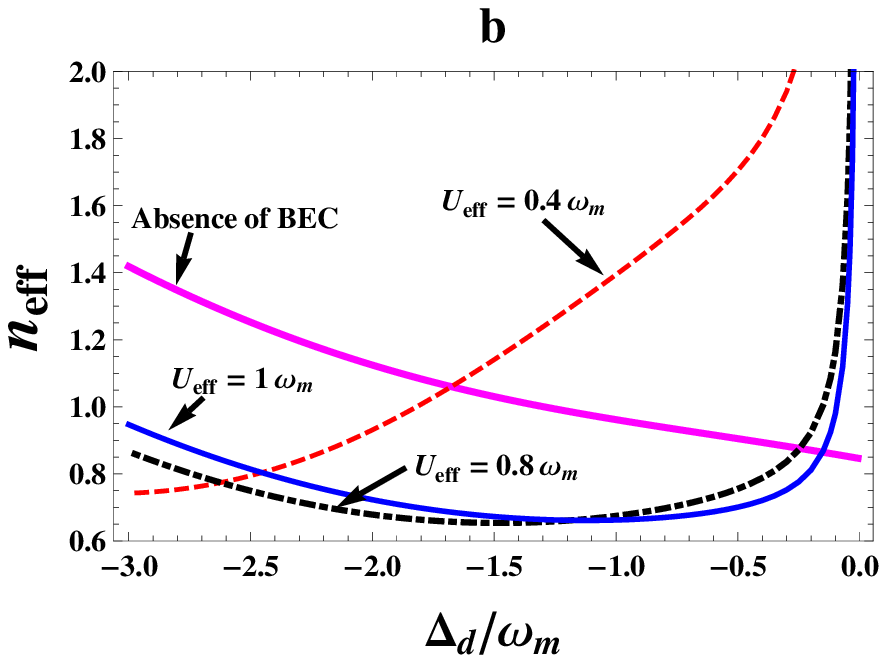}\\
\end{tabular}
\caption{(color online) Cold damping feedback scheme: (a) Plot of the mirror's position variances $\left\langle \delta q^{2} \right\rangle $ (solid line) and the mirror's momentum variances $\left\langle \delta p^{2} \right\rangle $ (dot dashed line) versus the normalized effective detuning ($\Delta_{d}/\omega_{m}$) in the absence of BEC and for three different values of atomic two-body interaction with $U_{eff}=0.4\omega_{m}$, $U_{eff}=0.8\omega_{m}$ and $U_{eff}=\omega_{m}$. Other parameters chosen are: $\Gamma_{m}=10^{-7}\omega_{m}$, $\kappa=5\omega_{m}$, $G=2\omega_{m}$, $\beta=0.05$, $\nu=\omega_{m}$, $g_{c}=\omega_{m}$, $\omega_{fb}=5\omega_{m}$, $g_{cd}=0.4$, $\lambda=0.85$ and $k_{B}T/\hbar\omega_{m}=10^{3}$. (b) Plot of effective phonon number with dimensionless effective detuning in the absence of BEC (thick Solid line) and for three different values of atomic two-body interaction with $U_{eff}=0.4\omega_{m}$ (dashed line), $U_{eff}=0.8\omega_{m}$ (thin Solid line) and $U_{eff}=\omega_{m}$ (dot dashed line). Other parameters used are same as in fig(a).}
\label{13}
\end{figure}

\begin{figure}[h]
\hspace{-0.0cm}
\includegraphics [scale=0.8]{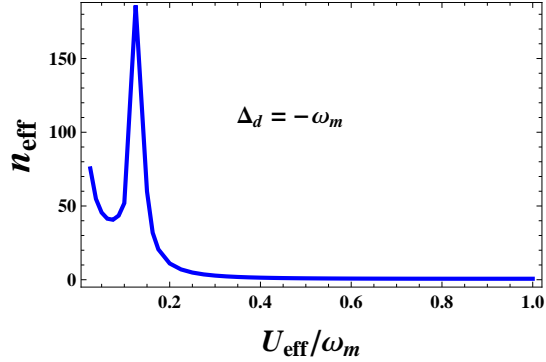}
\caption{(color online) The effective phonon number versus the normalized atomic two-body interaction at $\Delta_{d}=-\omega_{m}$ in cold damping feedback scheme. Rest of the parameters are same as figure 13(a).}
\label{14}
\end{figure}

Now, we characterize the steady state energy of the mirror in the cold damping feedback scheme. The optimal cooling conditions can be obtained using the stability conditions which are modified in this scheme given in the Appendix II. The oscillator variances represented by Eqns. (\ref{delq}) and (\ref{delp}) and the mean energy of the oscillator given by Eqn. (\ref{energy}), using the corresponding position spectrum for the cold damping case, are calculated  numerically with the help of MATHEMATICA 8.0. This shows that, also using cold damping, there is no equipartition of energy i.e., $\langle \delta q^{2} \rangle \neq \langle \delta p^{2} \rangle$ in general. Fig.13(a) illustrates both the variances $\langle \delta q^{2} \rangle$ and $\langle \delta p^{2} \rangle$ as a function of dimensionless detuning ($\Delta_{d}/\omega_{m}$) for the absence of BEC and three different values of condensate two-body interaction ($U_{eff}$), $U_{eff}=0.4\omega_{m}$, $U_{eff}=0.8\omega_{m}$ and $U_{eff}=1\omega_{m}$. As it can be seen from the figure, the displacement and momentum variances decrease significantly on adding BEC to the system. Moreover, both variances decrease with the increase in condensate two-body interaction. However, as $\Delta_{d}$ approaches zero, oscillator variances increase drastically in the presence of BEC. Therefore, one can infer that the presence of BEC helps in cooling the mirror for smaller detunings. So, in order to approach the optimal cooling conditions in cold damping scheme, the sum of these variances given by equations (\ref{delq}) and (\ref{delp}) is to be minimized which can be done by increasing the condensate two-body interaction. The effective phonon number given by Eqn. (\ref{neff}) is shown in Fig.13(b) as a function of dimensionless detuning ($\Delta_{d}/\omega_{m}$) for the absence of BEC (thick solid line) and the three different values of atom-atom interaction,  $U_{eff}=0.4\omega_{m}$(dashed line), $U_{eff}=0.8\omega_{m}$(thin solid line) and $U_{eff}=1\omega_{m}$(dot dashed line). From the figure, we observe that the minimum value of effective mean excitation number is reached for $U_{eff}=0.8\omega_{m}$ which is $0.658$ with $\Delta_{d}=-1.5\omega_{m}$. Although, as $\Delta_{d} \rightarrow 0$, the cavity field amplitude quadrature $\delta q_{b}(\omega)$ becomes insensitive to the mirror motion and $n_{eff}$ increases considerably with increasing $U_{eff}$. So, the ground state cooling can be approached by adding BEC to the system involving cold damping feedback scheme for lower detunings. Hence, by varying the atom-atom interaction, one can optimize the cooling process to acheive the ground state. Using cavity self-cooling scheme, least value of $n_{eff}$ is acheived in the absence of BEC while using cold damping feedback technique, minimum value is obtained by adding BEC to the system. Moreover, with BEC, we are getting the least value of effective phonon number using back-action cooling scheme. We have also shown the variation of effective phonon number $n_{eff}$ as a function of $U_{eff}/\omega_{m}$ in Fig.(14). It clearly illustrates the fact that, in the presence of BEC, the ground state cooling of the mechanical oscillator can be achieved for higher condensate two-body interaction only. This sudden increase in the value of $n_{eff}$ for very small atom-atom interaction is the consequence of the decrease in the effective mechanical damping rate of the mirror given by Eqn.(\ref{gcd}) in this region. Since in this chosen parameter regime, the second term in Eqn.(\ref{gcd}) becomes excessively small and even negative for very small $U_{eff}$. Moreover, we have also examined that by taking $\omega_{fb}=0.5\omega_{m}$, keeping other parameters same as before, the uncertainty principle is stratified for both displacement and momentum variances. This implies that $\left\langle \delta q^{2} \right\rangle \simeq \left\langle \delta p^{2} \right\rangle \simeq 1/2$, therefore, energy equipartiton is satisfied. However, the ground state could not be approached in this regime i.e., we are getting $n_{eff}>1$.

\section{Conclusion}

In this paper, we have studied how the back-action cooling and cold damping feedback schemes help in cooling the mirror to its quantum ground state by using a BEC confined in an optical cavity. The atom-atom two body interaction can be used as a new handle to cool the quantum device in both the schemes. It provides a systematic control of the system which can be altered either by using number of atoms or s-wave scattering length. Both the techniques show distinct displacement spectrums involving energy exchange between the different modes of the system. A coherent control over the feedback parameters, $g_{cd}$ and $\omega_{fb}$, can manipulate the displacement spectrum in the cold damping feedback scheme. In back-action cooling, the least value of effective phonon number is obtained without BEC whereas in cold damping, it is obtained with BEC. The effective temperature of the oscillator does not vary much for higher condensate two-body interaction with self-cooling. Also in this scheme, higher negative detuning gives lesser value of effective phonon number in the presence of BEC as compared to that in the absence of BEC. As we compare both the schemes, the minimum value of effective phonon number is obtained using cavity self-cooling with BEC. We have analyzed that both the techniques help in approaching the quantum ground state of the oscillator. Dicrimination in the ideal cooling conditions for these schemes exhibits that  back-action dynamics is more convenient in good cavity limit ($\kappa<<\omega_{m}$) while cold damping is preferable in bad cavity limit ($\kappa>>\omega_{m}$).

\section{Acknowledgements}

A. Bhattacherjee and Neha Aggarwal acknowledge financial support from the Department of Science and Technology, New Delhi for financial assistance vide grant SR/S2/LOP-0034/2010. Sonam Mahajan acknowledges University of Delhi for the University Teaching Assistantship.

\section{Appendix I}

Following correlations are satisfied by the input noise operators \citep{10,16,52,62} : 

\begin{equation}
\langle b_{in}(t) b_{in}(t')\rangle =\langle b_{in}^{\dagger}(t) b_{in}(t')\rangle =0,
\end{equation}

\begin{equation}
\langle b_{in}(t) b_{in}^{\dagger}(t')\rangle =\delta(t-t'),
\end{equation}

The noise operator due to the Brownian motion of the mirror is given as $W(t)=i \sqrt{\Gamma_{m}}[\xi_{m}^{\dagger}(t)-\xi_{m}(t)]$ which satisfies the following correlation \citep{52}: 

\begin{equation}
\langle W(t) W(t')\rangle = \frac{\Gamma_{m}}{\omega_{m}} \int \frac{d \omega}{2 \pi} e^{-i \omega (t-t')} \omega \left[1+\coth \left ( {\frac{\hbar \omega}{2 k_{B} T}} \right )\right],
\end{equation}

where $T$ is the finite temperature of the bath connected to the cantilever and $k_{B}$ is the Boltzmann constant. Brownian noise is the random thermal noise which arises from the stochastic motion of the mechanical oscillator (mirror). It should be noted that Brownian noise is non-Markovian in nature. The quantum Brownian motion of the movable mirror gives rise to the thermal noise term in the measured phase noise spectrum of the optical field reflected from the cavity \citep{10,16}.

The amplitude and phase quadratures of the input noise operator satisfy the following correlations in Fourier space \citep{62}:

$\langle q_{in}(\omega) q_{in}(\omega') \rangle=2 \pi \delta(\omega+\omega')$,
$\langle p_{in}(\omega) p_{in}(\omega') \rangle=2 \pi \delta(\omega+\omega')$,
$\langle q_{in}(\omega) p_{in}(\omega') \rangle=2i \pi \delta(\omega+\omega')$,
$\langle p_{in}(\omega) q_{in}(\omega') \rangle=-2i \pi \delta(\omega+\omega')$.

Also in fourier space, the correlation function for the brownian noise operator is given as \citep{62}:

$\langle W(\omega) W(\omega') \rangle=2 \pi \frac{\Gamma_{m}}{\omega_{m}} \omega \left[1+\coth \left ( {\frac{\hbar \omega}{2 k_{B} T}} \right )\right] \delta(\omega+\omega')$

The correlation of vacuum field quadrature in the Fourier space is
$\langle p_{v}(\omega) p_{v}(\omega') \rangle=2 \pi \delta(\omega+\omega').$ 

\section{Appendix II}

In the back-action cooling scheme, two non-trivial stability conditions for the system in the absence of BEC, obtained by applying the Routh-Hurwitz criterion, are given as:
\begin{equation}
S_{1}=a_{0}>0,
\end{equation}
\begin{equation}
S_{2}=(a_{3}a_{2}a_{1}-a_{4}a_{1}^{2}-a_{0}a_{3}^{2})>0,
\end{equation}
where 
\begin{equation}
a_{0}=\omega_{m}^{2}(\frac{\kappa^{2}}{4}+\Delta_{d}^{2})+\omega_{m}\beta^{2}G^{2}\Delta_{d},
\end{equation}
\begin{equation}
a_{1}=\Gamma_{m}(\Delta_{d}^{2}+\frac{\kappa^{2}}{4})+\omega_{m}^{2}\kappa,
\end{equation}
\begin{equation}
a_{2}=\frac{\kappa^{2}}{4}+\Delta_{d}^{2}+\Gamma_{m}\kappa+\omega_{m}^{2},
\end{equation}
\begin{equation}
a_{3}=\Gamma_{m}+\kappa,
\end{equation}
\begin{equation}
a_{4}=1.
\end{equation}
While the stability conditions, evaluated by applying the Routh-Hurwitz criterion, for the system in the presence of BEC are as follows:
\begin{equation}
S_{3}=b_{0}>0,
\end{equation}
\begin{equation}
S_{4}=(b_{5}b_{4}b_{3}+b_{6}b_{1}b_{5}-b_{6}b_{3}^{2}-b_{2}b_{5}^{2})>0,
\end{equation}
where
\begin{equation}
b_{0}=\frac{\beta_{1}\beta_{2}\kappa^{2}\omega_{m}^{2}}{4}+\omega_{m}^{2}\beta_{1}\beta_{2}\Delta_{d}^{2}+4G^{2} \beta^{2} \omega_{m} \Delta_{d} \beta_{1} \beta_{2}+4g_{c}^{2}\Delta_{d}\beta_{1}\omega_{m}^{2},
\end{equation}
\begin{equation}
b_{1}=\beta_{1}\beta_{2}\kappa \omega_{m}^{2}+\frac{\beta_{1}\beta_{2}\kappa^{2}\Gamma_{m}}{4}+\beta_{1} \beta_{2} \Gamma_{m} \Delta_{d}^{2}+4 g_{c}^{2} \Delta_{d} \beta_{1} \Gamma_{m},
\end{equation}
\begin{equation}
b_{2}=\frac{\kappa^{2}\omega_{m}^{2}}{4}+\omega_{m}^{2}\Delta_{d}^{2}+4G^{2} \beta^{2} \omega_{m} \Delta_{d}+\beta_{1} \beta_{2} \Gamma_{m} \kappa + \frac{\beta_{1}\beta_{2}\kappa^{2}}{4}+\beta_{1} \beta_{2} \Delta_{d}^{2}+ 4 g_{c}^{2} \Delta_{d} \beta_{1} + \beta_{1} \beta_{2} \omega_{m}^{2},
\end{equation}
\begin{equation}
b_{3}=\beta_{1}\beta_{2}(\Gamma_{m}+\kappa)+ \Gamma_{m}(\Delta_{d}^{2}+\frac{\kappa^{2}}{4})+ \kappa \omega_{m}^{2}, 
\end{equation}
\begin{equation}
b_{4}=\omega_{m}^{2}+ \Gamma_{m} \kappa+ \frac{\kappa^{2}}{4} + \Delta_{d}^{2}+ \beta_{1}\beta_{2},
\end{equation}
\begin{equation}
b_{5}=\Gamma_{m}+ \kappa, 
\end{equation}
\begin{equation}
b_{6}=1.
\end{equation}
Moreover, the Routh-Hurwitz criterion is equivalent to the condition by imposing all the poles of mechanical susceptibility ($\chi_{eff}^{cd}(\omega)$) in lower complex half plane. Hence, one gets a non-trivial modified stability condition for the cold damping case in the presence of BEC as follows:
\begin{equation}\label{s5}
S_{5}=(c_{6}c_{3}^{2}+c_{2}c_{5}^{2}-c_{5}c_{4}c_{3}-c_{6}c_{5}c_{1})>0,
\end{equation}
where
\begin{equation}
\begin{split}
c_{1}=-i(\Delta_{d}^{2}+\frac{\kappa^{2}}{4})(\omega_{fb}\omega_{m}^{2}+\omega_{fb}\beta_{1}\beta_{2}+\Gamma_{m}\beta_{1} \beta_{2})-i \beta_{1} \beta_{2}(\omega_{fb}\omega_{m}^{2}+ \Gamma_{m} \kappa \omega_{fb}+ \omega_{m}^{2}\kappa)\\-i 4 g_{c}^{2} \Delta_{d} \beta_{1}(\omega_{fb}+\Gamma_{m})-i 2G\beta\omega_{m}g_{cd}\beta_{1}\beta_{2}\lambda\omega_{fb}-i 4G^{2}\beta^{2}\Delta_{d}\omega_{fb}\omega_{m}, 
\end{split}
\end{equation}
\begin{equation}
\begin{split}
c_{2}=(\Delta_{d}^{2}+\frac{\kappa^{2}}{4})(\omega_{m}^{2}+ \Gamma_{m}\omega_{fb}+ \beta_{1}\beta_{2})+ \beta_{1} \beta_{2}(\omega_{m}^{2}+ \Gamma_{m} \omega_{fb}+ \omega_{fb} \kappa+ \Gamma_{m}\kappa)\\+ \omega_{m}^{2}\kappa \omega_{fb}+ 4g_{c}^{2} \Delta_{d}\beta_{1}+ G\beta \omega_{m} g_{cd} \lambda \omega_{fb}\kappa+ 4 G^{2}\beta^{2}\omega_{m}\Delta_{d}, 
\end{split}
\end{equation}
\begin{equation}
c_{3}=i\left[(\Delta_{d}^{2}+\frac{\kappa^{2}}{4})(\omega_{fb}+\Gamma_{m})+ \omega_{fb}(\omega_{m}^{2}+ \Gamma_{m} \kappa+ \beta_{1}\beta_{2})+ \kappa \omega_{m}^{2}+ \beta_{1}\beta_{2}(\kappa + \Gamma_{m})+ 2G\beta\omega_{m}g_{cd}\lambda\omega_{fb} \right], 
\end{equation}
\begin{equation}
c_{4}=-\left[(\Delta_{d}^{2}+\frac{\kappa^{2}}{4})+ \omega_{m}^{2} + \omega_{fb}(\kappa + \Gamma_{m})+ \beta_{1}\beta_{2}+ \kappa \Gamma_{m} \right], 
\end{equation}
\begin{equation}
c_{5}=-i\left[\omega_{fb}+ \kappa + \Gamma_{m} \right], 
\end{equation}
\begin{equation}\label{c6}
c_{6}=1.
\end{equation}
In the absence of BEC, the stability condition for the system involving cold damping feedback can be obtained by putting all the BEC parameters to be zero in Eqns. (\ref{s5})-(\ref{c6}).

\end{document}